\documentclass[aps,prd,onecolumn,notitlepage,eqsecnum,nofootinbib,floatfix,superscriptaddress]{revtex4-1}

\usepackage{amsmath, amsthm, amssymb, amsfonts, amsbsy,mathrsfs}

\usepackage{calligra,bm}
\usepackage{stmaryrd}

\usepackage[hidelinks,bookmarks=true]{hyperref} 
\hypersetup{pdfstartview=FitH,pdfhighlight=/O,colorlinks=false}

\begin{document}

\title{Volume average regularization for the Wheeler-DeWitt equation}

\author{Justin C. Feng}
\affiliation{Theory Group, Department of Physics, The University of Texas at Austin, Austin, 78712 Texas, USA}

\preprint{UTTG-04-18}


\begin{abstract}
In this article, I present a volume average regularization for the second functional derivative operator that appears in the metric-basis Wheeler-DeWitt equation. Naively, the second functional derivative operator in the Wheeler-DeWitt equation is infinite, since it contains terms with a factor of a delta function or derivatives of the delta function. More precisely, the second functional derivative contains terms that are only well defined as a distribution---these terms only yield meaningful results when they appear within an integral. The second functional derivative may, therefore, be regularized by performing an integral average of the distributional terms over some finite volume; I argue that such a regularization is appropriate if one regards quantum general relativity (from which the Wheeler-DeWitt equation may be derived) to be the low-energy effective field theory of a full theory of quantum gravity. I also show that a volume average regularization can be viewed as a natural generalization of the same-variable second partial derivative for an ordinary multivariable function. Using the regularized second functional derivative operator, I construct an approximate solution to the Wheeler-DeWitt equation in the low-curvature, long-distance limit.
\end{abstract}


\maketitle


\section{Introduction}
In quantum geometrodynamics,\footnote{In this article, quantum geometrodynamics refers specifically to the canonical formulation of quantum general relativity that uses the three-metric (or its inverse) as configuration space variables \cite{Kiefer2009}.} states may be described by a wave functional\footnote{Throughout this article, the symbol $g^{\cdot \cdot}$ in the arguments of functionals refers to the inverse metric; I do this to distinguish $g^{\cdot \cdot}$ from the determinant of the metric $g$, and to avoid any confusion with regard to indices.} $\Psi=\Psi[g^{\cdot \cdot}]$, which is a functional of a postive-definite inverse metric $g^{ij}=g^{ij}(y)$ for a three-manifold $\Sigma$, which I assume to be compact and without boundary. In quantum geometrodynamics, wave functionals satisfy the Wheeler-DeWitt equation \cite{DeWitt1967,Wheeler1968,Kiefer2009,Kiefer2012QG}, which takes the following form:\footnote{In some cases, one might wish to use some type of Laplace-Beltrami operator in place of the second functional derivative operator, either of the type briefly mentioned in \cite{DeWitt1967} or the type proposed in \cite{FengMatzner2017prd}. As discussed in \cite{FengMatzner2017prd}, the choice depends on the invariance principle required of the wavefunctional. Laplace-Beltrami operators are often used in minisuperspace models \cite{Kiefer2009,Kiefer2012QG}.  For the sake of simplicity, I follow \cite{DeWitt1967} and simply use the second functional derivative operator in the Wheeler-DeWitt equation.}
\begin{equation}\label{SFD-WdW}
\begin{aligned}
\hat{\mathcal{H}} \, \Psi =\hbar^2 \, {G}^{ijkl} \, \frac{\delta^2 \Psi}{\delta g^{ij}\delta g^{kl}} + \left( R -2 \, \Lambda \right) {\sqrt{g}} \, \Psi = 0 ,
\end{aligned}
\end{equation}
%

\noindent where $R$ is the Ricci curvature scalar for the three-manifold $\Sigma$, $\Lambda$ is the cosmological constant, and the quantity $G^{ijkl}$ is defined as
\begin{equation} \label{SFD-GMetric}
G^{ijkl}:=\frac{2 \, \kappa^2}{\sqrt{g}} \left (g^{ik} \, g^{jl} + g^{il} \, g^{jk} - g^{ij} \, g^{kl}\right) ,
\end{equation}

\noindent where $\kappa=8 \, \pi \, G$, with $G$ being Newton's constant. The Wheeler-DeWitt equation (\ref{SFD-WdW}) is supplemented by the following constraint, called the momentum constraint,
\begin{equation} \label{SFD-MomentumConstraint}
\begin{aligned}
g^{ik} \, \nabla_k \left(\frac{2 \, \kappa}{\sqrt{g}} \, \frac{\delta \Psi}{\delta g^{ij}}\right)=0 ,
\end{aligned}
\end{equation}

\noindent where $\nabla_k$ is the covariant derivative on $\Sigma$. In quantum geometrodynamics, the wavefunctional $\Psi=\Psi[g^{\cdot \cdot}]$ satisfies the Wheeler-DeWitt equation and the momentum constraint. The dynamical content of quantum geometrodynamics is provided by the Wheeler-DeWitt equation; as originally pointed out in \cite{Higgs1958}, the momentum constraint is simply the requirement that the wavefunctional $\Psi=\Psi[g^{\cdot \cdot}]$ be invariant under coordinate transformations on $\Sigma$. 

One difficulty\footnote{Another difficulty with quantum geometrodynamics concerns the precise definition of the inner product, which is formally defined as a functional integral over three-geometries. Further discussion of the inner product is beyond the scope of this article; I refer the reader to \cite{Woodard1993}, the general discussion found in \cite{Kiefer2009,Kiefer2012QG} and the references contained therein.} with the Wheeler-DeWitt equation is that, naively, second functional derivatives evaluated at the same spacetime point generally\footnote{As pointed out in \cite{DeWitt1967}, this is not always the case, as one can avoid this with certain double integrals over the manifold. However, it is generally the case if the functional contains  single integrals or double integrals with more than two factors of the field in the integrand.} contain terms with a singular factor of $\delta^3(0)$ \cite{Kiefer2012QG}, and/or terms containing a factor of the derivatives of the Dirac delta function (as I will show in this article).\footnote{One can define a second functional derivative at a point without delta functions \cite{Morrison1998}, but in that case, one trades delta functions for differential operators for functions $f(y)$ over $\Sigma$---in particular, the quantity $\hat{\mathcal{H}} \, \Psi$ itself becomes a linear differential operator acting on functions $f(y)$ over $\Sigma$. One must then find a functional $\Psi$ such that the equation $\hat{\mathcal{H}} \, \Psi \, f(y) = 0$ is satisfied for all functions $f(y)$. The Wheeler-DeWitt equation is no longer just a constraint for every point $y$, but it is now a constraint on the whole of some function space, namely the space of functions $f(y)$ on the manifold $\Sigma$. It presently is difficult for me to imagine how one might obtain nontrivial solutions to the Wheeler-DeWitt equation under such a strong constraint, so I will not pursue this approach in this article.}
For this reason, the second functional derivative in the full Wheeler-DeWitt equation is only defined a formal manner \cite{Esposito2011,*Esposito2012,FengMatzner2017prd}. The presence of delta functions and their derivatives is an indication that second functional derivatives only make sense as distributions; plainly speaking, second functional derivatives are only meaningful when they appear inside integrals. By itself, this is not problematic. What makes this problematic is that the Ricci scalar term in (\ref{SFD-WdW}) is a multiplicative operator, and its action on the wavefunctional will yield an ordinary (non distributional) function, so that the Wheeler-DeWitt equation (\ref{SFD-WdW}) naively states that a distributional quantity is equal to some nondistributional quantity. In this sense, the Wheeler-DeWitt equation, as written in Eq. (\ref{SFD-WdW}), is ill defined.

The origin of the singular quantity $\delta^3(0)$ in the second functional derivative operator comes from the fact that the naive second functional derivative is formally a function of two points $x$ and $y$, and contains terms with factors of $\delta^3(x-y)$. Since the second derivative operator in the Wheeler-DeWitt equation is evaluated at a single point, one may argue that the singular quantity $\delta^3(0)$ follows from short distance behavior (in particular the limit in which $x \rightarrow y$). It is well known that perturbative quantum general relativity contains ultraviolet divergences,\footnote{Though one can absorb one-loop divergences for pure gravity into the four-dimensional Gauss-Bonnet term via field redefinitions \cite{tHooftVeltman1974}, ultraviolet divergences appear at two-loop order in perturbation theory \cite{GoroffSagnotti1985,*GoroffSagnotti1986}.} so one might expect the short distance limit $x \rightarrow y$ to yield divergences.\footnote{Furthermore, the perturbative nonrenormalizability of quantum general relativity suggests that quantum general relativity, and by extension the Wheeler-DeWitt equation [as given in Eq. (\ref{SFD-WdW})], are incomplete; the renormalization of quantum general relativity requires an infinite number of counterterms in the action, which will generate additional terms in the Wheeler-DeWitt equation. For this reason, it is often argued that quantum general relativity cannot be a fundamental theory, but it has also been suggested that perturbation theory may not be generating the correct asymptotic series for quantum GR \cite{Woodard2009}, and that gravitational effects can somehow regulate the divergences of quantum field theory \cite{DeWitt1964,*Ishametal1971,*Ishametal1972,*Casadio2012} (though as argued in \cite{Woodard2009}, there is little hope that such a feature of quantum GR, if it exists, can be seen in perturbation theory).}

The modern view,\footnote{See \cite{Donoghue2017} and the references therein for an overview of quantum gravity as an effective field theory, and \cite{Burgess2004} for a more detailed review.} of course, is that quantum general relativity is the low-energy effective field theory of a full theory of quantum gravity. Since the Wheeler-DeWitt equation can be derived\footnote{Again, I emphasize the point, argued in \cite{FengMatzner2017prd}, that the precise form of the second functional derivative operator depends on the definition for the path integral measure.} in a formal manner from the path integral for quantum general relativity \cite{Leutwyler1964,*HartleHawking1983,*Halliwell1988,*HalliwellHartle1991,*Barvinsky1993a,FengMatzner2017prd}, one might expect some approximation to the Wheeler-DeWitt equation to be valid \cite{Carlip2017}. If one imagines quantum general relativity to be a low energy effective field theory, then the second functional derivative operator in the Wheeler-DeWitt equation must be regularized somehow. In particular, since the $\delta(0)$ singularity comes from a short-distance limit, effective field theory requires a regularization for the second functional derivative operator in (\ref{SFD-WdW}). 

An ad-hoc regularization for the Wheeler-DeWitt equation was originally proposed by Bryce DeWitt in \cite{DeWitt1967}, which is simply to set the singular quantities $\delta^3(0)$ to zero (this is done in dimensional regularization \cite{Hamber2009}); this is used to obtain a WKB approximation for the Wheeler-DeWitt equation \cite{Cenaloetal2003,Kiefer2012QG}. Lattice regularizations have also been studied in the literature, particularly those based on Regge discretizations--see \cite{Hamberetal2011,*Hamberetal2012,Hamberetal2013}. In this article, I describe a continuum regularization, which can be viewed as a natural generalization of the second-order same-variable partial derivative for an ordinary multivariable function. In particular, I perform a volume average of the second functional derivative, using integrals performed over the distributional part of the naive second functional derivative operator. Such a regularization is appropriate if one views the Wheeler-DeWitt equation as a description of a low energy gravitational effective field theory, as effective field theories are formed by integrating out high-energy modes of the field. Compared to DeWitt's regularization, the volume average regularization I present in this article has the advantage of providing a parameter that controls the regularization (the averaging volume), and I will briefly argue that a volume average regularization can in some sense be regarded as a generalization of DeWitt's regularization. I must make two things clear: first, while the methods presented in this article are motivated by effective field theory considerations, I do not establish a precise connection between the covariant methods of effective field theory and the volume average regularization presented in this article, which is formulated for a spatial hypersurface. Second, I make no claim with regard to the UV behavior of quantum geometrodynamics and the problem of nonrenormalizability for perturbative quantum gravity; my goal in this paper is to present a possible framework in which one can nonperturbatively investigate the low energy features of quantum gravity.

This article is organized as follows. First, I present a motivating example using ordinary second derivatives and Kronecker delta functions, and construct by analogy an expression for the second functional derivative operators of the type that appear in the Wheeler-DeWitt equation. The resulting expression is interpreted as an averaging of the second functional derivative operator over some volume, and its derivation makes it clear that it is a generalization of the same-variable second partial derivative. I then derive the Hessian for the volume functional and Einstein-Hilbert action. The Hessians are then used to construct an approximate solution for the regularized Wheeler-DeWitt equation in the low-curvature, long-distance limit. Finally, I examine a minisuperspace restriction of the approximate solution for three-sphere geometries.



\section{The Volume Average Regularization}
In this section, I motivate the volume average regularization for the second functional derivative of a functional $F[\varphi]$ evaluated at a single point $x$. In particular, I intend to motivate a regularized expression for the following quantity,
\begin{equation} \label{SFD-SecondFunctionalDerivative0}
\frac{\delta^2 F}{\delta \varphi^A_x \, \delta \varphi^B_x} ,
\end{equation}

\noindent where $\varphi^A_x=\varphi^A(x)$ is a function on a manifold $\mathcal{M}$ of volume $V_{\mathcal{M}}$ and coordinate label $x^i$. Instead of simply stating the result, I will attempt to motivate it by showing that the volume average regularization is a natural generalization of the same-variable second partial derivative of an ordinary multivariable function.

\subsection{The functional Hessian}
I begin by reviewing the definition of the second functional derivative (or the functional ``Hessian''). The second functional derivative is typically defined in terms of the Taylor expansion of the functional $F[\varphi]$:
\begin{equation} \label{SFD-TaylorExpansionFunctional}
\begin{aligned}
F[\varphi+\delta \varphi]:=&\, F[\varphi] + \sum_A \int_{\mathcal{M}} \, \frac{\delta F}{\delta \varphi^A_x}  \, \delta \varphi^A_x \, d^n x + \frac{1}{2!}\sum_{AB} \int_{\mathcal{M}} \int_{\mathcal{M}}  \, \frac{\delta^2 F}{\delta \varphi^A_x \,  \delta \varphi^B_y}  \, \delta \varphi^A_x \, \delta \varphi^B_y \> d^n x \, d^n y + \mathcal{O}(\delta \varphi^3) ,
\end{aligned}
\end{equation}

\noindent where I define the functions $\delta \varphi^A_x:=\delta \varphi^A(x)$ and $\delta \varphi^A_y:=\delta \varphi^A(y)$; if the manifold $\mathcal{M}$ has boundary $\partial \mathcal{M}$, I assume that the support of $\delta \varphi^A_x:=\delta \varphi^A(x)$ and $\delta \varphi^A_y:=\delta \varphi^A(y)$ does not reach a neighborhood of any point on the boundary $\partial \mathcal{M}$ (this way, I can neglect boundary terms). In this article, I do not employ summation convention for capital Latin indices $(A, B, ..., I, J, ...)$. Given the Taylor expansion (\ref{SFD-TaylorExpansionFunctional}), one can identify the second functional derivative, or the ``Hessian'' of the functional $F[\varphi]$:
\begin{equation} \label{SFD-SecondFunctionalDerivative1}
\begin{aligned}
\frac{\delta^2 F}{\delta \varphi^A_x \,  \delta \varphi^B_y}.
\end{aligned}
\end{equation}

\noindent Now consider a functional $F[\varphi]$ given by an integral of the form
\begin{equation} \label{SFD-LocalFunctional}
\begin{aligned}
F[\varphi]=\int_{\mathcal{M}}f(\varphi, x) \, \sqrt{g}\, d^n x .
\end{aligned}
\end{equation}

\noindent The Taylor expansion of $F[\varphi]$ will, in general, contain second-order terms of the form:
\begin{equation} \label{SFD-RegularTerms}
\begin{aligned}
\int_{\mathcal{M}} \sum_{AB} \left(\mathcal{F}_{AB} \,  \delta \varphi^A \, \delta \varphi^B \, \right)  \sqrt{g}\, d^nx .
\end{aligned}
\end{equation}

\noindent The above may be rewritten as
\begin{equation} \label{SFD-RegularTermsRewritten}
\begin{aligned}
\int_{\mathcal{M}} \int_{\mathcal{M}} \sum_{AB} \left(\{\mathcal{F}_{AB}\}_{(x,y)} \,  \delta \varphi^A_x \, \delta \varphi^B_y \,\right) \tilde{\delta}(x,y) \, \sqrt{g_x} \sqrt{g_y} \, d^n x \, d^n y ,
\end{aligned}
\end{equation}

\noindent where the brackets $\{ \>\> \}_{(x,y)}$ represent the symmetrization,
\begin{equation} \label{SFD-Symmetrization}
\begin{aligned}
\{ T^I \}_{(x,y)}:=\frac{1}{2} \left(T^I(x)+T^I(y)\right),
\end{aligned}
\end{equation}

\noindent and $\tilde{\delta}(x,y)$ is the covariant delta function, defined by the property,
\begin{equation} \label{SFD-CovariantDeltaFunction}
\begin{aligned}
\int_{\Sigma} \varphi(y) \, \tilde{\delta}(x,y) \, \sqrt{g_y} \> d^n y=\varphi(x) \>\>\>\>\> \Rightarrow \>\>\>\>\> \tilde{\delta}(x,y)= \frac{\delta^n(x-y)}{\sqrt{g_y}} ,
\end{aligned}
\end{equation}

\noindent with $\delta^n(x-y)$ being the $n$-dimensional Dirac delta function. Equation (\ref{SFD-RegularTermsRewritten}) indicates that in general, the second functional derivative of a functional, as defined by the Taylor expansion (\ref{SFD-TaylorExpansionFunctional}), contains terms with delta functions.

Now consider what happens if the functional depends on derivatives of $\varphi^A(x)$. For instance, consider the functional
\begin{equation} \label{SFD-LocalFunctionalWithDerivs}
\begin{aligned}
S[\varphi]=\int_{\mathcal{M}}\mathcal{L}(\varphi^A, \nabla_i \varphi^A, x) \, \sqrt{g} \, d^n x.
\end{aligned}
\end{equation}

\noindent In general, the Taylor expansion of $S[\varphi]$ to second order will contain terms of the following form:
\begin{equation} \label{SFD-DerivativeTerms}
\begin{aligned}
&\int_{\mathcal{M}} \int_{\mathcal{M}}\sum_{AB} \biggl( \{\mathcal{C}^{i}_{AB}\}_{(x,y)} \,  \delta \varphi^A_x \> ( \nabla_i^y \delta \varphi^B_y  ) \biggr) \tilde{\delta}(x,y) \, \sqrt{g_x} \sqrt{g_y} \, d^n x \, d^n y \\
&\int_{\mathcal{M}} \int_{\mathcal{M}} \sum_{AB} \biggl(\{\mathcal{D}^{ij}_{AB}\}_{(x,y)} \, (\nabla_i^x {\delta \varphi^A_x}) \, (\nabla_j^y {\delta \varphi^B_y})\biggr) \tilde{\delta}(x,y) \, \sqrt{g_x} \sqrt{g_y} \, d^n x \, d^n y ,
\end{aligned}
\end{equation}

\noindent where $\nabla_i^x$ is the covariant derivative taken with respect to $x$ and $\nabla_i^y$ is the covariant derivative taken with respect to $y$. It is possible to add boundary terms to convert the above expressions to an integral of the form (\ref{SFD-RegularTerms}):
\begin{equation} \label{SFD-RegularForm}
\begin{aligned}
\int_{\mathcal{M}} \sum_{AB} \left(\mathcal{C}_{AB} \,  \delta \varphi^A \, \delta \varphi^B \, \right)  \sqrt{g}\, d^nx .
\end{aligned}
\end{equation}

\noindent However, in doing so, one will encounter terms containing both delta functions and derivatives of delta functions. Naively setting $x=y$ will yield a divergent result.

Of course, the reader may be well aware that delta functions and their derivatives are not really functions in the usual sense---they are distributions and are only meaningful if they appear once inside an integral. Recalling that the covariant delta function $\tilde{\delta}(x,y)$ is defined by the property (\ref{SFD-CovariantDeltaFunction}), I may use the divergence theorem to assign a definition for the covariant derivative of the delta function $\tilde{\delta}(x,y)$,
\begin{equation} \label{SFD-DerivativeDistributions}
\begin{aligned}
\int_{\mathcal{M}}\sum_{AB} v^i(y) \, \nabla_i^y \tilde{\delta}(x,y) \, \sqrt{g_y}\> d^n y &=- \int_{\mathcal{M}} \int_{\mathcal{M}}\sum_{AB} \nabla_i^y v^i(y) \,  \tilde{\delta}(x,y) \, \sqrt{g_y}\> d^n y =-\nabla_i^x v^i(x) ,
\end{aligned}
\end{equation}

\noindent for some vector field $v^i(x)$.

\subsection{Ordinary second derivatives: A motivating example}
To motivate the regularized expression for the same-point second functional derivative, I consider an example for ordinary multivariable functions. I examine Hessian of a function $f(x)$ of quantities $x^I$:
\begin{equation}\label{SFD-MechDeriv}
\begin{aligned}
\frac{\partial^2 f}{\partial x^I \, \partial x^J} .
\end{aligned}
\end{equation}

\noindent Now suppose that the Hessian takes the form
\begin{equation}\label{SFD-MechDerivForm}
\begin{aligned}
\frac{\partial^2 f}{\partial x^I \, \partial x^J}=\Phi_{I J}(x) \, \delta_{I J}  + \Omega_{IJ}(x),
\end{aligned}
\end{equation}

\noindent where $\delta_{ I J }$ is the Kronecker delta, which is the discrete-value analog of the Dirac delta function $\delta^n(y-z)$.\footnote{Compare the expression $\sum_J \, x^J \, \delta_{IJ}=x^I$ with its integral counterpart $\int_{M} f(z) \, \delta^n (y-z) \, d^n z=f(y)$.} Again, I must remind the reader that in this article, no sum is implied over repeated capital Latin indices. If I simply set $I=J$, I obtain the second derivative for a single value of the index $I$:
\begin{equation}\label{SFD-MechDerivII}
\begin{aligned}
\frac{\partial^2 f}{\partial x^I \, \partial x^I}=\Phi_{I I}(x) + \Omega_{II}(x) .
\end{aligned}
\end{equation}

\noindent Now suppose that, for some reason, I want to obtain an expression for $\frac{\partial^2 f}{\partial x^I \, \partial x^I}$ without explicitly setting $I=J$. If I set $\Omega_{IJ}(x)=0$, I may do this by performing the following sum:
\begin{equation}\label{SFD-MechDerivII-2}
\begin{aligned}
\sum_{J} \, \frac{\partial^2 f}{\partial x^I \, \partial x^J}=\sum_{J} \, \Phi_{I J}(x) \, \delta_{I J}=\Phi_{I I}(x) .
\end{aligned}
\end{equation}

\noindent Note that for $\Omega_{IJ}(x)=0$, (\ref{SFD-MechDerivII-2}) and (\ref{SFD-MechDerivII}) yield the same result. However, for $\Omega_{IJ}(x) \neq 0$, the sum in (\ref{SFD-MechDerivII-2}) does not yield (\ref{SFD-MechDerivII}). To recover (\ref{SFD-MechDerivII}) in the case where $\Omega_{IJ}(x) \neq 0$, I decompose the second derivative (\ref{SFD-MechDerivForm}) into a part proportional to the Kronecker delta, which I call $\text{D}\llbracket \cdot \rrbracket$, and a part that does not contain any factor of the Kronecker delta, which I call $\text{N}\llbracket \cdot \rrbracket$. For (\ref{SFD-MechDerivForm}), I have
\begin{equation}\label{SFD-MechDerivFormDecomp}
\begin{aligned}
\text{D}\left\llbracket\frac{\partial^2 f}{\partial x^I \, \partial x^J}\right\rrbracket &=\Phi_{I J}(x) \, \delta_{I J} \\
\text{N}\left\llbracket\frac{\partial^2 f}{\partial x^I \, \partial x^J}\right\rrbracket &=\Omega_{IJ}(x).
\end{aligned}
\end{equation}

\noindent With this decomposition, I construct the following:
\begin{equation}\label{SFD-MechDerivII-3}
\begin{aligned}
\frac{\partial^2 f}{\partial x^I \, \partial x^I}=\left(\sum_{J} \, \text{D}\left\llbracket\frac{\partial^2 f}{\partial x^I \, \partial x^J}\right\rrbracket \right) + \left. \text{N}\left\llbracket\frac{\partial^2 f}{\partial x^I \, \partial x^J}\right\rrbracket \right|_{I=J} .
\end{aligned}
\end{equation}

\noindent It is straightforward to verify that the above construction (\ref{SFD-MechDerivII-3}) yields the same result as (\ref{SFD-MechDerivII}).

\subsection{Second functional derivatives at a single point}
The generalization of Eq. (\ref{SFD-MechDerivII-3}) to second functional derivatives comes from identifying the Dirac delta function $\delta^n(y-z)$ as the continuous-index analog of the Kronecker delta $\delta_{IJ}$ and the integral over $\mathcal{M}$ as the continuous-index analog of the sum. Suppose I have a quantity $H^{AB}=H^{AB}(x,y)$ such that its transformation under coordinate transformations on $\mathcal{M}$ leaves the following integral unchanged:\footnote{If the indices $(A,B)$ are formed from the indices of the coordinate basis $(i,j)$, then I require that $H^{AB}$ transforms as a tensor.}
\begin{equation} \label{SFD-SecondFunctionalDerivative2}
\int_{\mathcal{M}} \int_{\mathcal{M}} H^{AB}(x,y) \, \frac{\delta^2 F}{\delta \varphi^A_x \, \delta \varphi^B_y} \, d^n x \, d^n y .
\end{equation}

\noindent I split the second functional derivative into a distributional part and a nondistributional part:
\begin{equation} \label{SFD-SecondFunctionalDerivative3}
\frac{\delta^2 F}{\delta \varphi^A_x \, \delta \varphi^B_y}= \text{DS}\left\llbracket\frac{\delta^2 F}{\delta \varphi^A_x \, \delta \varphi^B_y}\right\rrbracket + \text{ND}\left\llbracket\frac{\delta^2 F}{\delta \varphi^A_x \, \delta \varphi^B_y}\right\rrbracket ,
\end{equation}

\noindent where the distributional part $\text{DS}\llbracket \cdot \rrbracket$ is the part of a quantity containing a factor of a delta function $\tilde{\delta}(y,z)$ or its derivatives, and the nondistributional $\text{ND}\llbracket \cdot \rrbracket$ is the part of the second functional derivative that does not contain delta functions $\tilde{\delta}(y,z)$ or its derivatives.

By analogy to (\ref{SFD-MechDerivII-3}), I construct the following regularization for the second functional derivative (with the equivalence relation $\cong$ indicating the regularized expression),
\begin{equation} \label{SFD-SecondFunctionalDerivativeSamePointDefinition}
\begin{aligned}
 H^{AB}(x) \, \frac{\delta^2 F}{\delta \varphi^A_x \, \delta \varphi^B_x}
 \cong \, & \frac{\sqrt{g_x}}{V} \int_{\mathcal{M}} \left( H^{AB}{(x,y)} \> \> \text{DS} \left\llbracket\frac{\delta^2 F}{\delta \varphi^A_x \, \delta \varphi^B_y} \right\rrbracket \right) d^n y  + H^{AB}(x) \left.  \text{ND}\left\llbracket\frac{\delta^2 F}{\delta \varphi^A_x \, \delta \varphi^B_y}\right\rrbracket \right|_{y=x} ,
\end{aligned}
\end{equation}

\noindent where $H_{AB}(x):=H_{AB}(x,x)$ and $V$ is a volume parameter. The inverse volume factor of $1/V$ in front of the first term must be included so that Eq. (\ref{SFD-SecondFunctionalDerivativeSamePointDefinition}) is dimensionally correct; $\text{DS}\llbracket \cdot \rrbracket$ has the same units as its argument, and one must compensate for the volume element $d^ny$ with a factor of $1/V$. The factor of $\sqrt{g_y}$ in front of the first term is put in so that the first term satisfies the same transformation properties as the second term; the second functional derivative contains a factor of $\sqrt{g_x} \, \sqrt{g_y}$  (also note that the covariant delta function $\tilde{\delta}(x,y)$ eliminates a factor of $\sqrt{g_y}$ in the integral). One might recognize the integral in the first term of (\ref{SFD-SecondFunctionalDerivativeSamePointDefinition}) as an average of the second functional derivative over some volume $V$.

\subsection{The regularized Wheeler-DeWitt equation}\label{Sec-SFDOPWDW}
Equation (\ref{SFD-SecondFunctionalDerivativeSamePointDefinition}) suggests the following regularization for the second derivative operator in the Wheeler-DeWitt equation,
\begin{equation}\label{SFD-WdWHamReg}
\begin{aligned}
\tilde{G}^{abmn}(y)\frac{\delta^2 \Psi}{\delta g^{ab}_y \,  \delta g^{mn}_y} \cong \frac{\sqrt{g_y}}{V} \,\int_{\Sigma} \left(\{ \tilde{G}^{abmn} \}_{(y,z)} \, \text{DS}\biggl\llbracket \frac{\delta^2 \Psi}{\delta g^{ab}_y \,  \delta g^{mn}_z} \biggr\rrbracket \right) d^3 z + \tilde{G}^{abmn}(y) \left.\text{ND}\biggl\llbracket \frac{\delta^2 \Psi}{\delta g^{ab}_y \,  \delta g^{mn}_z} \biggr\rrbracket\right|_{z=y} ,
\end{aligned}
\end{equation}

\noindent where $\tilde{G}^{abmn}$ is the following tensor, constructed from $G^{abmn}$ (\ref{SFD-GMetric}):
\begin{equation} \label{SFD-GMetric2}
\begin{aligned}
\tilde{G}^{abmn}&=\frac{\sqrt{g}}{2 \, \kappa^2} \> G^{abmn}=g^{ab} \, g^{mn} - g^{am} \, g^{bn} - g^{an} \, g^{bm} .
\end{aligned}
\end{equation}

\noindent The regularized Wheeler-DeWitt equation is then
\begin{equation}\label{SFD-WdWReg}
\begin{aligned}
\frac{2 \, \hbar^2 \, \kappa^2}{V} \,\int_{\Sigma} \left(\{ \tilde{G}^{abmn} \}_{(y,z)} \, \text{DS}\biggl\llbracket \frac{\delta^2 \Psi}{\delta g^{ab}_y \,  \delta g^{mn}_z} \biggr\rrbracket \right) d^3 z + \hbar^2 \, G^{abmn}(y) \left.\text{ND}\biggl\llbracket \frac{\delta^2 \Psi}{\delta g^{ab}_y \,  \delta g^{mn}_z} \biggr\rrbracket\right|_{z=y} + \left( R(y) -2 \, \Lambda \right) \sqrt{g_y} \> \Psi =0.
\end{aligned}
\end{equation}

\noindent In the limit $V \rightarrow 0$, the above expression diverges, as one might expect---as discussed earlier, the second functional derivative operator in the Wheeler-DeWitt equation is formally divergent, since it is naively the limit of a distributionally valued quantity.

For compact three-manifolds $\Sigma$ with finite volume $V_{\Sigma}[g^{\cdot \cdot}]$, it is tempting (one might even say that it is ``natural'') to choose $V=V_{\Sigma}[g^{\cdot \cdot}]$ in equation (\ref{SFD-SecondFunctionalDerivativeSamePointDefinition}). For manifolds with infinite volume $V_{\Sigma}[g^{\cdot \cdot}] \rightarrow \infty$, the first term in (\ref{SFD-SecondFunctionalDerivativeSamePointDefinition}) vanishes; this is the sense in which a volume averaging regularization can be viewed as a generalization of DeWitt's \textit{ad hoc} regularization \cite{DeWitt1967}: $\delta^3(0)=0$. One might imagine formulating a model for quantum gravity with the replacement (by fiat) of the second functional derivative by the expression (\ref{SFD-WdWHamReg}) where $V=V_{\Sigma}[g^{\cdot \cdot}]$; in this case, the distributional part of (\ref{SFD-SecondFunctionalDerivativeSamePointDefinition}) is nonvanishing for small volumes, but vanishes in the large-volume limit. Assuming certain properties\footnote{In particular, one assumes $\Psi[g^{\cdot \cdot}]=\exp(S[g^{\cdot \cdot}])$, where $S[g^{\cdot \cdot}]$ is a local functional of $g^{ij}$ (by local, I mean that $S[g^{\cdot \cdot}]$ can be written as an integral over $\Sigma$ with an integrand that depends only on $g^{ij}$ and its derivatives at a single point).} for the wavefunctional, one recovers the Einstein-Hamilton-Jacobi equation in the large-volume limit, irrespective of the value for $\hbar$; this behavior suggests a possible mechanism in which this $V=V_{\Sigma}[g^{\cdot \cdot}]$ quantum gravity model ``classicalizes'' in the large-volume limit. 

On the other hand, if one imagines quantum geometrodynamics to be the result of some low-energy gravitational effective field theory, then it may be appropriate to perform a volume averaging that corresponds to integrating out short distance degrees of freedom. In the context of effective field theory, it is appropriate to choose $V=v_0$, where $v_0$ is a fixed, finite volume determined by the length scale corresponding to the high frequency modes that have been integrated out in the effective field theory; for quantum gravity, it is natural to choose $v_0$ to be the Planck volume $(\hbar \, \kappa)^{3/2}$. Since effective field theory provides a clear physical justification for the choice $V=v_0$ (the physical justification for $V=V_\Sigma[g^{\cdot \cdot}]$ is less clear to me at present), I shall focus on the effective field theory viewpoint and the choice $V=v_0$ for the remainder of this article.


\section{Wavefunctionals and Hessians of Invariant Integrals}
\subsection{Wavefunctionals and the chain rule}
In this section, I will derive expressions for the second functional derivative of the volume functional and the Einstein-Hilbert functional. This is a long section, and the calculations are tedious, so I wish to first provide some motivation for deriving the second functional derivative of these functionals. Recall the momentum constraint (\ref{SFD-MomentumConstraint}), which I rewrite here:
\begin{equation} \label{SFD-MomentumConstraint2}
\begin{aligned}
g^{ik} \, \nabla_k \left(\frac{2 \, \kappa}{\sqrt{g}} \, \frac{\delta \Psi}{\delta g^{ij}}\right)=0 .
\end{aligned}
\end{equation}

\noindent If the functional $\Psi[g^{\cdot \cdot}]$ is constructed from integrals over a compact three-manifold $\Sigma$, then the momentum constraint (\ref{SFD-MomentumConstraint2}) implies that the integrals must be invariant under coordinate transformations \cite{Higgs1958}. The integrals themselves must be constructed out of curvature invariants, since they are the only scalar quantities that can be constructed from the three-metric \cite{Hamberetal2011,*Hamberetal2012,Hamberetal2013}. Under the assumption that any covariant multiple integral constructed from the three-metric can be expanded in terms of products of single integrals of a curvature invariant, it follows that the wavefunctional can be written as a function of (single) integrals of curvature invariants. If the three-manifold has finite volume, the wavefunctional will also depend on the volume functional of the manifold:
\begin{equation}\label{SFD-VolumeFunctional}
V_{\Sigma}[g^{\cdot \cdot}]=\int_{\Sigma} \sqrt{g} \, d^3 y .
\end{equation}

The simplest nontrivial curvature invariant is the Ricci scalar $R$, and its integral is the (three-dimensional) Einstein-Hilbert action:
\begin{equation}\label{SFD-3EH}
S_{EH}[g^{\cdot \cdot}]=\int_{\Sigma} \, R \, \sqrt{g} \, d^3 y .
\end{equation}

\noindent A simple ansatz for the wavefunctional is one in which the wavefunctional is a function of the following functional:
\begin{equation}\label{SFD-Slambda}
S_{\lambda}:=S_{EH}[g^{\cdot \cdot}]- 2\, \lambda \, V_{\Sigma}[g^{\cdot \cdot}] =\int_{\Sigma} (R -2\,\lambda)\, \sqrt{g} \, d^3 y .
\end{equation}

\noindent In particular, I write:
\begin{equation}\label{SFD-WavefunctionEH}
\begin{aligned}
\Psi[g^{\cdot \cdot}]=\Psi(S_{\lambda}[g^{\cdot \cdot}]) .
\end{aligned}
\end{equation}

\noindent I now perform the variation of the wavefunctional:
\begin{equation}\label{EHWF-WavefunctionEHVar}
\begin{aligned}
\Delta \Psi[g^{\cdot \cdot}]:&=\Psi[g^{\cdot \cdot}+\delta g^{\cdot \cdot}]-\Psi[g^{\cdot \cdot}]\\
&=\Psi(S[g^{ij}]+\Delta S)-\Psi(S[g^{ij}])\\
&=\frac{\partial \Psi}{\partial S} \, \Delta S+\frac{1}{2}\frac{\partial^2 \Psi}{\partial S^2} \Delta S^2 + \mathcal{O}(\Delta S^3),
\end{aligned}
\end{equation}

\noindent where $\Delta S:=S[g^{ij}+\delta g^{ij}]-S[g^{ij}]$. Upon performing a Taylor expansion of $\Delta S$ in $\delta g^{ij}$ to second order [cf., Eq. (\ref{SFD-TaylorExpansionFunctional})], Eq. (\ref{EHWF-WavefunctionEHVar}) becomes
\begin{equation}\label{SFD-WavefunctionEHVar1}
\begin{aligned}
\Delta \Psi[g^{\cdot \cdot}]
&=\frac{\partial \Psi}{\partial  S_{\lambda}} \, \delta  S_{\lambda} + \frac{1}{2}\frac{\partial \Psi}{\partial S_{\lambda}} \, \int_{\Sigma}  \int_{\Sigma} \, \frac{\delta^2 S_{\lambda}}{\delta g^{ab}_y \,  \delta g^{mn}_z}  \, \delta g^{ab}_y \, \delta g^{mn}_z \, d^3y \, d^3 z +\frac{1}{2}\frac{\partial^2 \Psi}{\partial  S_{\lambda}^2} \delta  S_{\lambda}^2+\mathcal{O}([\delta g^{\cdot \cdot}]^3) ,
\end{aligned}
\end{equation}

\noindent where $\delta g_y^{ab}=\delta g^{ab}(y)$ and $\delta g_z^{mn}=\delta g^{mn}(z)$. The variation $\delta  S_{\lambda}$ can be written in terms of a functional derivative,
\begin{equation}\label{SFD-EHFirstVar}
\begin{aligned}
\delta  S_{\lambda}&= \int_{\Sigma}\, \frac{\delta  S_{\lambda}}{\delta g^{ab}} d^3 y ,
\end{aligned}
\end{equation}

\noindent and $\delta  S_{\lambda}^2$ may be written as
\begin{equation}\label{SFD-EHFirstVarSquared}
\begin{aligned}
\delta  S_{\lambda}^2&= \int_{\Sigma} \int_{\Sigma} \, \frac{\delta  S_{\lambda}}{\delta g^{ab}_y} \, \frac{\delta  S_{\lambda}}{\delta g^{mn}_z} d^3 y \, d^3 z .
\end{aligned}
\end{equation}

\noindent The variation of the functional derivative (\ref{SFD-WavefunctionEHVar1}) may then be rewritten as:
\begin{equation}\label{SFD-WavefunctionEHVar2}
\begin{aligned}
\Delta \Psi[g^{\cdot \cdot}]
=\frac{\partial \Psi}{\partial  S_{\lambda}} \, \delta  S_{\lambda} + \frac{1}{2}\int_{\Sigma} \int_{\Sigma} \, \biggl[&\frac{\partial \Psi}{\partial  S_{\lambda}} \,  \frac{\delta^2  S_{\lambda}}{\delta g^{ab}_y \,  \delta g^{mn}_z} + \frac{\partial^2 \Psi}{\partial  S_{\lambda}^2} \, \frac{\delta  S_{\lambda}}{\delta g^{ab}_y} \, \frac{\delta  S_{\lambda}}{\delta g^{mn}_z} \,  \biggl]\, \delta g^{ab}_y \, \delta g^{mn}_z \> d^3 y \, d^3 z .
\end{aligned}
\end{equation}

\noindent From the above expression, I identify the second functional derivative:
\begin{equation}\label{SFD-WavefunctionFuncDeriv}
\begin{aligned}
\frac{\delta^2 \Psi}{\delta g^{ab}_y \,  \delta g^{mn}_z} &=\frac{\partial \Psi}{\partial S_{\lambda}} \,  \frac{\delta^2  S_{\lambda}}{\delta g^{ab}_y \,  \delta g^{mn}_z} + \frac{\partial^2 \Psi}{\partial S^2} \, \frac{\delta  S_{\lambda}}{\delta g^{ab}_y} \, \frac{\delta  S_{\lambda}}{\delta g^{mn}_z} .
\end{aligned}
\end{equation} 

\noindent The above expression depends on the second functional derivative (the functional ``Hessian'') of the functional $S_{\lambda}[g^{\cdot \cdot}]$; in the remainder of this section, I will derive expressions for the second functional derivative of $S_{\lambda}$.

\subsection{The Hessian of the volume functional}
First, I compute the second functional derivative (the functional Hessian) for the volume functional $V_{\Sigma}[g^{\cdot \cdot}]$. I first work out a few useful expressions. Since I am expanding to second order in $\delta g^{ij}$, it does not suffice to work in terms of the first order variation $\delta g_{ij} = - g_{ia} \, g_{jb} \, \delta g^{ab}$ for the metric. In general, what is needed is the second order expression for the change in the metric $\Delta g_{ij}$:
\begin{equation} \label{SFD-MetricVariation}
\begin{aligned}
\Delta g_{ij} = - g_{ia} \, g_{jb} \, \delta g^{a b}  +  g_{ai} \, g_{jm} \, g_{bn} \, \delta g^{ab} \, \delta g^{mn} + \mathcal{O}([\delta g^{\cdot \cdot}]^3) .
\end{aligned}
\end{equation}

\noindent The second order expression follows from the property $g^{ij} \, g_{j n}=\delta^i_n$; in particular, Eq. (\ref{SFD-MetricVariation}) follows from demanding that the following expression holds to second order:
\begin{equation} \label{SFD-MetricInverseIdentity}
(g^{ab} + \delta g^{ab}) (g_{b j}+\Delta g_{b j})=\delta^a_j .
\end{equation}

\noindent The property $g^{ij} \, g_{j n}=\delta^i_n$ may also be used to derive the following expressions for the derivatives of $g^{ij}$ and $g_{ij}$:
\begin{equation} \label{SFD-InverseDerivative1}
\frac{\partial g_{mn}}{\partial s} = - g_{mi} \, g_{nj} \frac{\partial g^{ij}}{\partial s}
\end{equation}
\begin{equation} \label{SFD-InverseDerivative2}
 g^{mn} \frac{\partial g_{mn}}{\partial s} = - g_{mn}\frac{\partial g^{mn}}{\partial s}.
\end{equation}

\noindent Using the Jacobi determinant formula with (\ref{SFD-InverseDerivative2}) I obtain the following result:
\begin{equation} \label{SFD-VolumeFactorDerivative} 
\frac{\partial \sqrt{g}}{\partial s}= \frac{1}{2 \, \sqrt{g}} \frac{\partial g}{\partial s}= \frac{1}{2} \, \sqrt{g} \, g^{ij} \, \frac{\partial g_{ij}}{\partial s}=-\frac{1}{2} \, \sqrt{g} \, g_{ij} \, \frac{\partial g^{ij}}{\partial s}.
\end{equation}

\noindent Another result is the following:
\begin{equation} \label{SFD-MetricPartialDerivComponents}
\frac{\partial g^{ij}}{\partial g^{mn}}=\frac{1}{2} \left(\delta^i_m \, \delta^j_n + \delta^i_n \, \delta^j_m\right).
\end{equation}

\noindent Using (\ref{SFD-InverseDerivative1}), (\ref{SFD-VolumeFactorDerivative}), and (\ref{SFD-MetricPartialDerivComponents}), I compute the change in the volume element, keeping terms to second order:
\begin{equation} \label{SFD-VolumeElementVariationASO}
\begin{aligned}
\Delta \sqrt{g} 
    &:=\left(\frac{\partial \sqrt{g}}{\partial g^{mn}}\right)\delta g^{mn}+ \frac{1}{2} \left(\frac{\partial^2 \sqrt{g} }{\partial g^{ab}\,\partial g^{mn}}\right) \delta g^{ab} \, \delta g^{mn} + \mathcal{O}([\delta g^{\cdot \cdot}]^3)\\
& \, =-\frac{1}{2} \sqrt{g} \> g_{ab} \>  \delta g^{a b} + \frac{1}{2} \sqrt{g}  \, Y_{abmn} \, \delta g^{ab} \, \delta g^{mn} + \mathcal{O}([\delta g^{\cdot \cdot}]^3)
\end{aligned}
\end{equation}

\noindent where $Y_{abmn}$ is defined as
\begin{equation}\label{SFD-Yabmn}
\begin{aligned}
Y_{abmn}:=\frac{1}{4} \, \left(g_{ab} \,  g_{mn} + g_{ma} \, g_{nb} + g_{mb} \, g_{na}\right) .
\end{aligned}
\end{equation}

The change in the volume functional may be written as
\begin{equation}\label{SFD-VolumeFunctionalChange}
\begin{aligned}
\Delta V_{\Sigma} &= V_{\Sigma}[g^{\cdot \cdot}+\delta g^{\cdot \cdot} ]-V_{\Sigma}[g^{\cdot \cdot}]=\int_{\Sigma} \Delta \sqrt{g} \, d^3 y.
\end{aligned}
\end{equation}

\noindent Using the result (\ref{SFD-VolumeElementVariationASO}) and inserting a delta function $\delta(y,z)$ into the integral, I obtain
\begin{equation}\label{SFD-VolumeFunctionalChange2}
\begin{aligned}
\Delta V_{\Sigma} &=-\frac{1}{2} \int_{\Sigma} \, g_{ab} \>  \delta g^{a b}\, \sqrt{g} \, d^3 y + \frac{1}{2} \int_{\Sigma} \, Y_{abmn} \, \delta g^{ab} \, \delta g^{mn} \, \sqrt{g}  \, d^3 y + \mathcal{O}([\delta g^{\cdot \cdot}]^3) \\
&=-\frac{1}{2} \int_{\Sigma} \, g_{ab} \>  \delta g^{a b}\, \sqrt{g} \, d^3 y + \frac{1}{2} \int_{\Sigma} \, \left\{Y_{abmn}\right\}_{y,z} \, \delta g^{ab}_y \, \delta g^{mn}_z \, \tilde{\delta}(y,z) \, \sqrt{g_y} \sqrt{g_z}  \, d^3 y \, d^3 z + \mathcal{O}([\delta g^{\cdot \cdot}]^3).
\end{aligned}
\end{equation}

\noindent I can read off the first and second functional derivatives from the above by comparing it with the functional Taylor expansion (\ref{SFD-TaylorExpansionFunctional}):
\begin{equation}\label{SFD-VolumeFunctionalDerivative}
\frac{\delta V_{\Sigma}}{\delta g^{ab}} = - \frac{1}{2} g_{ab}\, \sqrt{g},
\end{equation}
\begin{equation}\label{SFD-VolumeFunctionalHessian}
\frac{\delta^2 V_{\Sigma}}{\delta g^{ab}_y \delta g^{mn}_z} = \left\{Y_{abmn}\right\}_{y,z} \, \tilde{\delta}(y,z) \sqrt{g_y} \, \sqrt{g_z}.
\end{equation}

\subsection{The Hessian of the Einstein-Hilbert action}
I now compute the second functional derivative (the functional Hessian) for the Einstein-Hilbert action $S_{EH}[g^{\cdot \cdot}]$ itself. Expressions for the Hessian of the Einstein-Hilbert action do appear in the literature (particularly in work which makes use of the saddle-point approximation for quantum gravity---see for instance \cite{Grossetal1982,Codelloetal2009}). I present for the benefit of the reader an explicit derivation of the Hessian. For the remainder of this article, I assume that the manifold $\Sigma$ is compact and without boundary.

I begin by writing down an expression for the change in the Ricci scalar. Though it may be strange to do so before performing variations of the Christoffel symbols, the variation of the Christoffel symbols is rather complicated at second order (later, I show that the first-order expressions for the Christoffel symbols suffice). If I obtain a variation in the Ricci scalar first, I can identify the places where second-order terms in the variation of the Christoffel symbols are needed, if at all. In fact, I show that the second variation of the Einstein-Hilbert action does involve second-order variations in the Christoffel symbols.

The change in the Ricci curvature is worked out in the Appendix [see Eq. (\ref{A-RicciTensorVariation})],
\begin{equation}\label{SFD-RicciTensorVariation}
\Delta {R}_{ab} = \nabla_i \Delta {\Gamma}^i_{ba}-\nabla_b \Delta {\Gamma}^i_{ia}+\Delta {\Gamma}^i_{is} \>\Delta {\Gamma}^s_{ba}- \Delta {\Gamma}^i_{bs} \>\Delta {\Gamma}^s_{ia} ,
\end{equation}

\noindent where $\Delta {\Gamma}^i_{jk}$ [Eq. (\ref{A-ConnectionTransformed})] is the change in the Christoffel symbols. In terms of $\Delta {R}_{ab}$ and $\delta g^{ab}$, the change in the Ricci scalar is
\begin{equation}\label{SFD-RicciScalarVariation}
\begin{aligned}
\Delta R &= \delta g^{ab} \,  {R}_{ab} + g^{ab} \, \Delta {R}_{ab} + \delta g^{ab} \, \Delta {R}_{ab} .
\end{aligned}
\end{equation}

\noindent One can combine equations (\ref{SFD-VolumeElementVariationASO}), (\ref{SFD-RicciTensorVariation}), and (\ref{SFD-RicciScalarVariation}) to obtain the following expression for the variation of the Einstein-Hilbert action [see Appendix for the algebra leading up to Eq. (\ref{A-3EHVar2c})]:
\begin{equation}\label{SFD-3EHVar2}
\begin{aligned}
\Delta S_{EH} & := S_{EH}[g^{\cdot \cdot} + \delta g^{\cdot \cdot}] - S_{EH}[g^{\cdot \cdot}]\\
&= \int_{\Sigma} \left( \Delta R \, \sqrt{g} + R \, \Delta \sqrt{g} + \Delta R \, \Delta \sqrt{g}\right) d^3 y\\
& = \delta S_{EH} + \int_{\Sigma} \biggl[ g^{ab} \, (\nabla_i \Delta {\Gamma}^i_{ba}-\nabla_b \Delta {\Gamma}^i_{ia} ) + \frac{1}{2}\left( R \, Y_{abmn} - {R}_{ab} \,g_{mn} \right) \delta g^{ab} \, \delta g^{mn} + g^{ab} \left( \Delta {\Gamma}^i_{is} \>\Delta {\Gamma}^s_{ba} - \Delta {\Gamma}^i_{bs} \>\Delta {\Gamma}^s_{ia} \right)\\
& \>\>\>\>\>\>\>\>\>\>\>\>\>\>\>\>\>\>\>\>\>\>\>\>\>\>\>\>\>\>
 + \left(\delta g^{ab} -  \frac{1}{2} \, g^{ab} \, g_{mn} \, \delta g^{mn} \right)(\nabla_i \Delta {\Gamma}^i_{ba}-\nabla_b \Delta {\Gamma}^i_{ia})    + \mathcal{O}([\delta g^{\cdot \cdot}]^3)\biggr] \sqrt{g} \, d^3 y ,
\end{aligned}
\end{equation}

\noindent where $\delta S_{EH}$ is the first-order variation of the action given by
\begin{equation}\label{SFD-3EH1Var}
\begin{aligned}
\delta S_{EH}:=\int_{\Sigma} \, \left({R}_{ab} - \frac{1}{2}  \> g_{ab} \> R \right) \delta g^{ab} \, \sqrt{g} \,  d^3 y .
\end{aligned}
\end{equation}

\noindent I recognize that the term $g^{ab} \, (\nabla_i \Delta {\Gamma}^i_{ba}-\nabla_b \Delta {\Gamma}^i_{ia} )$ in (\ref{SFD-3EHVar2}) is a total divergence--it is a boundary term. Since the the manifold $\Sigma$ is assumed to be compact and without boundary, I eliminate this boundary term. The variation of the Einstein-Hilbert action becomes (\ref{A-3EHVar4C})
\begin{equation}\label{SFD-3EHVar3}
\begin{aligned}
\Delta S_{EH} = \delta S_{EH} + \int_{\Sigma} \biggl[& g^{ab} \left( \Delta {\Gamma}^i_{is} \>\Delta {\Gamma}^s_{ba} - \Delta {\Gamma}^i_{bs} \>\Delta {\Gamma}^s_{ia} \right)  + \left(\delta g^{ab}  -  \frac{1}{2} \, g^{ab} \, g_{mn} \, \delta g^{mn} \right) (\nabla_i \Delta {\Gamma}^i_{ba}-\nabla_b \Delta {\Gamma}^i_{ia}) \\
&  + \frac{1}{2} \left( R \, Y_{abmn}  -   {R}_{ab} \,g_{mn}\right) \delta g^{ab} \> \delta g^{mn}  + \mathcal{O}([\delta g^{\cdot \cdot}]^3) \biggr] \sqrt{g} \> d^3 y ,
\end{aligned}
\end{equation}

\noindent Note that each time $\Delta {\Gamma}^i_{ab}$ appears in the above expression, it is either accompanied by a factor of $\delta g^{ab}$ or another factor of $\Delta {\Gamma}^i_{ab}$. It follows that only the first-order part of $\Delta {\Gamma}^i_{ab}$ contributes to second-order terms in (\ref{SFD-3EHVar3}). To obtain an expression for $\Delta S$ that is second order in the variations of the inverse metric $\delta g^{ab}$, it suffices to use an expression for $\Delta {\Gamma}^i_{ab}$ to first order in $\delta g^{ab}$. Recalling the definition of the Christoffel symbol,
\begin{equation} \label{SFD-ChristoffelSymbolB}
\Gamma^a_{ij}=\frac{1}{2} g^{ak} (\partial_i g_{kj}+\partial_j g_{ik}-\partial_k g_{ij}),
\end{equation}

\noindent it is not difficult to show that to first order, the variation of the Christoffel symbol takes the covariant form
\begin{equation} \label{SFD-ChristoffelSymbolVariationA}
\begin{aligned}
\delta \Gamma^a_{ij}=\frac{1}{2}g^{ak}(\nabla_i \Delta g_{kj}+\nabla_j \Delta g_{ik}-\nabla_k \Delta g_{ij}) + \mathcal{O}([\delta g^{\cdot \cdot}]^2) ,
\end{aligned}
\end{equation}

\noindent where $\Delta g_{ij}$ is defined in (\ref{SFD-MetricVariation}). To first order, one may use (\ref{SFD-MetricVariation}) to rewrite Eq. (\ref{SFD-ChristoffelSymbolVariationA}) in terms of variations of the inverse metric:
\begin{equation} \label{SFD-ChristoffelSymbolVariationB}
\begin{aligned}
\delta \Gamma^a_{ij} = \frac{1}{2}(g_{mi} \, g_{nj} \, g^{ak}\,  \nabla_k \delta g^{mn} - g_{nj} \, \nabla_i \delta g^{an} - g_{mi} \, \nabla_j \delta g^{ma}) + \mathcal{O}([\delta g^{\cdot \cdot}]^2) .
\end{aligned}
\end{equation}

\noindent Given (\ref{SFD-ChristoffelSymbolVariationB}), I may then rewrite the variation of the Einstein-Hilbert action (\ref{SFD-3EHVar3}) in terms of the first-order expressions $\delta \Gamma^a_{ij}$,
\begin{equation}\label{SFD-3EHVar4}
\begin{aligned}
\Delta S_{EH} = \delta S_{EH} + \int_{\Sigma} \biggl[& g^{ab} \, \delta {\Gamma}^i_{is} \>\delta {\Gamma}^s_{ba} - g^{ab} \, \delta {\Gamma}^i_{bs} \>\delta {\Gamma}^s_{ia} + \left(\delta g^{ab}  -  \frac{1}{2} \, g^{ab} \, g_{mn} \, \delta g^{mn} \right) (\nabla_i \delta {\Gamma}^i_{ba}-\nabla_b \delta {\Gamma}^i_{ia}) \\
&  + \frac{1}{2} \left( R \, Y_{abmn}  -   {R}_{ab} \,g_{mn}\right) \delta g^{ab} \> \delta g^{mn}  + \mathcal{O}([\delta g^{\cdot \cdot}]^3) \biggr] \sqrt{g} \> d^3 y ,
\end{aligned}
\end{equation}

\noindent where $\delta \Gamma^a_{ij}$ is given by (\ref{SFD-ChristoffelSymbolVariationB}). After an application of the divergence theorem, equation (\ref{SFD-3EHVar4}) for the variation of the Einstein-Hilbert action takes the following form [Eq. (\ref{A-3EHVar4C})],
\begin{equation}\label{SFD-3EHVar2C}
\begin{aligned}
\Delta S_{EH} &= \delta S_{EH} + \int_{\Sigma} \biggl[ Z_{abmn}^{ij} \, \nabla_i \delta g^{ab} \, \nabla_j \delta g^{mn}  + \frac{1}{2} \left( R \, Y_{abmn}  -   {R}_{ab} \,g_{mn}\right) \delta g^{ab} \> \delta g^{mn}  + \mathcal{O}([\delta g^{\cdot \cdot}]^3) \biggr] \sqrt{g} \> d^3 y,
\end{aligned}
\end{equation}

\noindent where $Z_{abmn}^{ij}$ is a tensor formed from terms containing products of $g_{ij}$, $g^{ij}$, and $\delta^i_j$. From metric compatibility, it follows that $Z_{abmn}^{ij}$ satisfies the property
\begin{equation}\label{SFD-ZTensorDeriv}
\nabla_k Z_{abmn}^{ij}=0 .
\end{equation}

\noindent Ultimately, the explicit form for $Z_{abmn}^{ij}$ is not important for the results in this paper; what matters is that it satisfies the property (\ref{SFD-ZTensorDeriv}). Nevertheless, I have derived the following explicit expression for $Z_{abmn}^{ij}$ in the Appendix [Eq. (\ref{A-ZTensor})]:
\begin{equation} \label{SFD-ScalarCZ}
\begin{aligned}
Z_{abmn}^{ij} &:=\frac{1}{4}\biggl(4\, g_{na} \, \delta_m^i \, \delta_b^j - 2 \, g_{mb} \, \delta_a^j \, \delta_n^i - g_{mb} \, g_{an}  \, g^{ij} + g^{ij} \, g_{mn} \, g_{ab} - 2 \, g_{mn} \, \delta^i_b \, \delta^j_a \biggr) .
\end{aligned}
\end{equation}

\noindent Equation (\ref{SFD-3EHVar2C}) may be converted into the following multiple integral:
\begin{equation}\label{SFD-3EHVar2D}
\begin{aligned}
\Delta S_{EH} \approx \delta S_{EH} + \int_{\Sigma}\int_{\Sigma} \biggl[& \{ Z_{abmn}^{ij} \}_{(y,z)} \, \nabla_i^y \delta g^{ab}_y \, \nabla_j^z \delta g^{mn}_z + \frac{1}{2} \{ R \, Y_{abmn}  -  {R}_{ab} \,g_{mn}\}_{(y,z)} \, \delta g^{ab}_y \, \delta g^{mn}_z \biggr] \tilde{\delta}(y,z) \sqrt{g_y} \sqrt{g_z} \, d^3 y \, d^3 z,
\end{aligned}
\end{equation}

\noindent where $\nabla_k^y$ and $\nabla_k^z$, respectively, denote covariant derivatives taken with respect to $y^i$ and $z^i$, $\delta g^{ab}_y:=\delta g^{ab}(y)$, and $\delta g^{mn}_z:=\delta g^{mn}(z)$. Recall (\ref{SFD-Symmetrization}), where the brackets $\{ \> \}_{(y,z)}$ denote the operation
\begin{equation}\label{SFD-Symmetrization2}
\begin{aligned}
\{ T^I \}_{(y,z)}=\frac{1}{2}\left(T^I(y) + T^I(z)\right) ,
\end{aligned}
\end{equation}

\noindent for some tensor $T^I=T^I(y)$.  $\tilde{\delta}(y,z)$ is the covariant three-dimensional delta function, defined by the property
\begin{equation}\label{SFD-CovDelta}
\begin{aligned}
\int_{\Sigma} \varphi(z) \, \tilde{\delta}(y,z) \, \sqrt{g_z} \> d^3 z=\varphi(z) \>\>\>\>\> \Rightarrow \>\>\>\>\> \tilde{\delta}(y,z)= \frac{\delta^3(y-z)}{\sqrt{g_z}} ,
\end{aligned}
\end{equation}

\noindent where $\varphi(z)$ is a scalar and $\delta^n(y-z)$ is the \textit{n}-dimensional Dirac delta function.

Applying the divergence theorem, I obtain
\begin{equation}\label{SFD-3EHVar2C2}
\begin{aligned}
\Delta S_{EH} \approx \delta S_{EH} + \int_{\Sigma}\int_{\Sigma} \biggl[& \{ Z_{abmn}^{ij} \}_{(y,z)} \, \left( \nabla_i^y \nabla_j^z\tilde{\delta}(y,z) \right)  \delta g^{ab}_y \,  \delta g^{mn}_z  \\
& + \frac{1}{2} \{ R \, Y_{abmn}  -   {R}_{ab} \,g_{mn} \}_{(y,z)} \, \delta g^{ab}_y \> \delta g^{mn}_z \, \tilde{\delta}(y,z) \biggr] \sqrt{g_y} \sqrt{g_z} \> d^3 y  \> d^3 z ,
\end{aligned}
\end{equation}

\noindent where I have used the fact that $\nabla_k Z_{abmn}^{ij}=0$, since $Z_{abmn}^{ij}$ is constructed from Kronecker deltas and the metric $g_{mn}$. By comparison with Eq. (\ref{SFD-TaylorExpansionFunctional}), I may write down the following expression for the second functional derivative (Hessian) as:
\begin{equation}\label{SFD-FuncDeriv}
\begin{aligned}
\frac{\delta^2 S_{EH}}{\delta g^{ab}_y \,  \delta g^{mn}_z}= 2\biggl[\{ Z_{abmn}^{ij} \}_{(y,z)} \, \left( \nabla_i^y \nabla_j^z\tilde{\delta}(y,z) \right) + \frac{1}{2} \{ R \, Y_{abmn}  -   {R}_{ab} \,g_{mn} \}_{(y,z)} \, \tilde{\delta}(y,z) \biggr] \sqrt{g_y} \sqrt{g_z}.
\end{aligned}
\end{equation}


\section{An Approximate Solution to the Wheeler-DeWitt Equation}
In this section, I obtain approximate solutions to the Wheeler-DeWitt equation, using the results obtained in the preceding sections.
\subsection{Second functional derivatives of the wavefunctional}
I now compute the regularized operator [Eq. (\ref{SFD-WdWHamReg})]:
\begin{equation}\label{SFD-WdWHamReg2}
\begin{aligned}
\tilde{G}^{abmn}(y)\frac{\delta^2 \Psi}{\delta g^{ab}_y \,  \delta g^{mn}_y} \cong \frac{\sqrt{g_y}}{V} \,\int_{\Sigma} \left(\{ \tilde{G}^{abmn} \}_{(y,z)} \, \text{DS}\biggl\llbracket \frac{\delta^2 \Psi}{\delta g^{ab}_y \,  \delta g^{mn}_z} \biggr\rrbracket \right) d^3 z + \tilde{G}^{abmn}(y) \left.\text{ND}\biggl\llbracket \frac{\delta^2 \Psi}{\delta g^{ab}_y \,  \delta g^{mn}_z} \biggr\rrbracket\right|_{z=y} .
\end{aligned}
\end{equation}

\noindent for wavefunctionals of the form $\Psi=\Psi(S_{\lambda})$. I now recall Eq. (\ref{SFD-WavefunctionFuncDeriv}):
\begin{equation}\label{SFD-WavefunctionFuncDeriv2}
\begin{aligned}
\frac{\delta^2 \Psi}{\delta g^{ab}_y \,  \delta g^{mn}_z} &=\frac{\partial \Psi}{\partial S_{\lambda}} \,  \frac{\delta^2 S_{\lambda}}{\delta g^{ab}_y \,  \delta g^{mn}_z} + \frac{\partial^2 \Psi}{\partial S_{\lambda}^2} \, \frac{\delta S_{\lambda}}{\delta g^{ab}_y} \, \frac{\delta S_{\lambda}}{\delta g^{mn}_z}.
\end{aligned}
\end{equation} 

\noindent From Eqs. (\ref{SFD-VolumeFunctionalDerivative}) and (\ref{SFD-3EH1Var}), the first functional derivative of $S_{\lambda}=S_{EH} - 2\, \lambda \, V_{\Sigma}$ (\ref{SFD-Slambda}) is
\begin{equation}\label{SFD-SlambdaFunctionalDeriv}
\frac{\delta S_{\lambda}}{\delta g^{ab}} = \left({R}_{ab} - \frac{1}{2}  \> g_{ab} \> \tilde{R} \right) \sqrt{g} ,
\end{equation}

\noindent where I have defined
\begin{equation}\label{SFD-TildeR}
\tilde{R}:=R - 2 \, \lambda .
\end{equation}

\noindent From Eq. (\ref{SFD-SlambdaFunctionalDeriv}), one can infer that the second term in (\ref{SFD-WavefunctionFuncDeriv2}) is nondistributional. Using (\ref{SFD-VolumeFunctionalHessian}) and (\ref{SFD-FuncDeriv}), one can construct the Hessian of $S_{\lambda}=S_{EH} - 2 \, \lambda \, V_{\Sigma}$:
\begin{equation}\label{SFD-SlambdaHessian}
\frac{\delta^2 S_{\lambda}}{\delta g^{ab}_y \,  \delta g^{mn}_z}= \biggl[2\{ Z_{abmn}^{ij} \}_{(y,z)} \, \left( \nabla_i^y \nabla_j^z\tilde{\delta}(y,z) \right) + \{ (R- \lambda/2) \, Y_{abmn}  -   {R}_{ab} \,g_{mn} \}_{(y,z)} \, \tilde{\delta}(y,z) \biggr] \sqrt{g_y} \sqrt{g_z}.
\end{equation}

\noindent Every term in the Hessian (\ref{SFD-SlambdaHessian}) contains a factor of the delta function or its derivatives. I can now identify the distributional part of the second functional derivative:
\begin{equation}\label{SFD-WavefunctionFuncDerivDS}
\begin{aligned}
\text{DS}\biggl\llbracket \frac{\delta^2 \Psi}{\delta g^{ab}_y \,  \delta g^{mn}_z} \biggr\rrbracket & = \frac{\partial \Psi}{\partial S_{\lambda}} \,  \frac{\delta^2 S_{\lambda}}{\delta g^{ab}_y \,  \delta g^{mn}_z} \\
& =\frac{\partial \Psi}{\partial S_{\lambda}} \, \biggl[2\{ Z_{abmn}^{ij} \}_{(y,z)} \left( \nabla_i^y \nabla_j^z\tilde{\delta}(y,z) \right)  + \{ (R - \lambda/2) \, Y_{abmn}  -   {R}_{ab} \,g_{mn} \}_{(y,z)} \, \tilde{\delta}(y,z) \biggr] \sqrt{g_y} \sqrt{g_z} . 
\end{aligned}
\end{equation}

\noindent To work out the explicit expression for (\ref{SFD-WavefunctionFuncDerivDS}), I begin by constructing the following integral:
\begin{equation}\label{SFD-FuncDerivInt}
\begin{aligned}
\int_{\Sigma} \left( \{ \tilde{G}^{abmn} \}_{(y,z)} \, \frac{\delta^2 S_{\lambda}}{\delta g^{ab}_y \,  \delta g^{mn}_z} \right) d^3 z&= \int_{\Sigma}\biggl[2\{ Z_{abmn}^{ij} \, \tilde{G}^{abmn}\}_{(y,z)} \, \left( \nabla_i^y \nabla_j^z\tilde{\delta}(y,z) \right)\biggr] \sqrt{g_y} \sqrt{g_z} \, d^3 z \\
&\>\>\>\>\> + \int_{\Sigma}  \biggl[ \{ \tilde{G}^{abmn} \, ( (R - \lambda/2) \, Y_{abmn}  -   {R}_{ab} \,g_{mn} )\}_{(y,z)} \, \tilde{\delta}(y,z) \biggr] \sqrt{g_y} \sqrt{g_z} \, d^3 z .
\end{aligned}
\end{equation}

\noindent It is straightforward to derive the following result for the two quantities $\delta=\delta(y,z)$ and $Q^{ij}=Q^{ij}(y,z)$:
\begin{equation}\label{SFD-DivergenceResult}
\begin{aligned}
\nabla_i^y (Q^{ij} \, \nabla_j^z \delta)-\nabla_j^z (\delta \, \nabla_i^y Q^{ij}) 
&= Q^{ij} \, \nabla_i^y \nabla_j^z \delta  - \delta \, \nabla_j^z \nabla_i^y Q^{ij} .
\end{aligned}
\end{equation}

\noindent Using the above result, I may rewrite (\ref{SFD-FuncDerivInt}) as
\begin{equation}\label{SFD-FuncDerivInt2}
\begin{aligned}
\int_{\Sigma} \left( \{ \tilde{G}^{abmn} \}_{(y,z)} \, \frac{\delta^2 S_{\lambda}}{\delta g^{ab}_y \,  \delta g^{mn}_z} \right) d^3 z&= \int_{\Sigma}\biggl[2\nabla_i^y \nabla_j^z \{ Z_{abmn}^{ij} \, \tilde{G}^{abmn}  \}_{(y,z)} \, \biggr] \tilde{\delta}(y,z) \sqrt{g_y}\sqrt{g_z} \, d^3 z \\
&\>\>\>\>\> + \int_{\Sigma}  \biggl[ \{ \tilde{G}^{abmn} \,((R -2/\lambda) \, Y_{abmn}  -   {R}_{ab} \,g_{mn} )\}_{(y,z)} \, \tilde{\delta}(y,z) \biggr] \sqrt{g_y}  \sqrt{g_z}\, d^3 z ,
\end{aligned}
\end{equation}

\noindent which becomes
\begin{equation}\label{SFD-FuncDerivInt3}
\begin{aligned}
\int_{\Sigma} \left( \{ \tilde{G}^{abmn} \}_{(y,z)} \, \frac{\delta^2 S_{\lambda}}{\delta g^{ab}_y \,  \delta g^{mn}_z} \right) d^3 z
&= 2 \biggl[\nabla_i^y \nabla_j^z \{ Z_{abmn}^{ij} \, \tilde{G}^{abmn}  \}_{(y,z)} \, \biggr]_{z=y} \, \sqrt{g} +   \tilde{G}^{abmn} \left( ( R-2/\lambda) \, Y_{abmn}  -   {R}_{ab} \,g_{mn}  \right)\sqrt{g}.
\end{aligned}
\end{equation}

\noindent Note that $Z_{abmn}^{ij}$ (\ref{SFD-ScalarCZ}) is a quadratic expression in $g_{ij}$ and $\delta^i_j$, and $\tilde{G}^{abmn}$ is a quadratic expression in $g_{ij}$. It follows that covariant derivatives of $Z_{abmn}^{ij} \, \tilde{G}^{abmn}$ vanishes by virtue of metric compatibility [$\nabla_k g_{ij}=0$ and $\nabla_k g^{ij}=0$; cf. (\ref{SFD-ZTensorDeriv})]. From the definition of $Y_{abmn}$ (\ref{SFD-Yabmn}):
\begin{equation}\label{SFD-FuncDerivInt4}
\begin{aligned}
\int_{\Sigma} \left( \{ \tilde{G}^{abmn} \}_{(y,z)} \, \frac{\delta^2 S_{\lambda}}{\delta g^{ab}_y \,  \delta g^{mn}_z} \right) d^3 z
&= \tilde{G}^{abmn} \, \biggl[ \frac{1}{4} (R-\lambda/2) ( g_{ab} \,  g_{mn} + 2 \, g_{am} \, g_{bn} )  -  {R}_{ab} \,g_{mn}  \biggr]\sqrt{g}.
\end{aligned}
\end{equation}

\noindent Using the definition (\ref{SFD-GMetric2}) for $\tilde{G}^{abmn}$, I work out the following quantities:
\begin{equation} \label{SFD-Linear}
\begin{aligned}
&\tilde{G}^{abmn} \, {g}_{am} \, {g}_{bn} = -9\\
&\tilde{G}^{abmn} \, {g}_{ab} \, {g}_{mn} = 3\\
&\tilde{G}^{abmn} \, {R}_{ab} \, {g}_{mn} =\tilde{G}^{abmn}\> g_{ab} \>  {R}_{mn}  = R\\
&\tilde{G}^{abmn} \,  {R}_{mn} \, {R}_{ab} = {R}^2 - 2 \, {R}^{mn} \, {R}_{mn} .
\end{aligned}
\end{equation}

\noindent I then use Eqs. (\ref{SFD-FuncDerivInt4}) and (\ref{SFD-Linear}) to obtain the following expression:
\begin{equation}\label{SFD-FuncDerivInt3c}
\begin{aligned}
\int_{\Sigma} \left( \{ \tilde{G}^{abmn} \}_{(y,z)}  \, \text{DS}\biggl\llbracket \frac{\delta^2 \Psi}{\delta g^{ab}_y \,  \delta g^{mn}_z} \biggr\rrbracket \right) d^3 z & = \frac{\partial \Psi}{\partial S_{\lambda}} \, \int_{\Sigma} \left( \{ \tilde{G}^{abmn} \}_{(y,z)} \, \frac{\delta^2 S_{EH}}{\delta g^{ab}_y \,  \delta g^{mn}_z} \right) d^3 z\\
&=  \frac{\partial \Psi}{\partial S_{\lambda}} \, \tilde{G}^{abmn} \, \biggl( \frac{1}{4} \left(R - \frac{\lambda}{2} \right) ( g_{ab} \,  g_{mn} + 2 \, g_{am} \, g_{bn} )  -  {R}_{ab} \,g_{mn}  \biggr)\sqrt{g}\\
&=-\frac{1}{8}\frac{\partial \Psi}{\partial S_{\lambda}} \,  \left( 38 \, R - 15 \, \lambda \right)\sqrt{g}.
\end{aligned}
\end{equation}

I now work out the nondistributional part of the second functional derivative of $\Psi$ in Eq. (\ref{SFD-WavefunctionFuncDerivND}). It is not too difficult to show that
\begin{equation}\label{SFD-WavefunctionFuncDerivdSdS}
\begin{aligned}
\frac{\delta S_{\lambda}}{\delta g^{ab}_y} \, \frac{\delta S_{\lambda}}{\delta g^{mn}_z}&= \left({R}_{ab}^y - \frac{1}{2}  \> g_{ab}^y \> \tilde{R}^y \right) \left({R}_{mn}^z - \frac{1}{2}  \> g_{mn}^z \> \tilde{R}^z \right) \sqrt{g_y} \,  \sqrt{g_z}\\
&= \biggl(
{R}_{ab}^y \, {R}_{mn}^z - \frac{1}{2}  \tilde{R}^z \, g_{mn}^z \, {R}_{ab}^y - \frac{1}{2} \tilde{R}^y \, g_{ab}^y \, {R}_{mn}^z + \frac{1}{4} \, g_{ab}^y  \,  g_{mn}^z \, \tilde{R}^y \, \tilde{R}^z\biggr) \sqrt{g_y} \,  \sqrt{g_z} ,
\end{aligned}
\end{equation}

\noindent again, recalling the definition $\tilde{R}:=R - 2 \, \lambda$ [Eq. (\ref{SFD-TildeR})]. Since Eq. (\ref{SFD-WavefunctionFuncDerivdSdS}) contains no delta functions, I write down the nondistributional part of the second functional derivative of $\Psi$:
\begin{equation}\label{SFD-WavefunctionFuncDerivND}
\begin{aligned}
\text{ND}\biggl\llbracket \frac{\delta^2 \Psi}{\delta g^{ab}_y \,  \delta g^{mn}_z} \biggr\rrbracket &= \frac{\partial^2 \Psi}{\partial S^2_{\lambda}} \, \frac{\delta S_{\lambda}}{\delta g^{ab}_y} \, \frac{\delta S_{\lambda}}{\delta g^{mn}_z}\\
&= \frac{\partial^2 \Psi}{\partial S^2_{\lambda}} \, \biggl( {R}_{ab}^y \, {R}_{mn}^z - \frac{1}{2}  \tilde{R}^z \, g_{mn}^z \, {R}_{ab}^y - \frac{1}{2} \tilde{R}^y \, g_{ab}^y \, {R}_{mn}^z + \frac{1}{4} \, g_{ab}^y  \,  g_{mn}^z \, \tilde{R}^y \, \tilde{R}^z \biggr) \sqrt{g_y} \,  \sqrt{g_z} .
\end{aligned}
\end{equation}

\noindent Making use of the symmetry in $\tilde{G}^{abmn}$ [recall that it is symmetric in the indices $(a, b)$ and $(m,n)$], Eq. (\ref{SFD-WavefunctionFuncDerivND}) yields the result:
\begin{equation}\label{SFD-WavefunctionFuncDerivND1B}
\begin{aligned}
\tilde{G}^{abmn}(y) \left.\text{ND}\biggl\llbracket \frac{\delta^2 \Psi}{\delta g^{ab}_y \,  \delta g^{mn}_z} \biggr\rrbracket\right|_{z=y}
= \tilde{G}^{abmn} \, \frac{\partial^2 \Psi}{\partial S^2_{\lambda}} \biggl( & {R}_{ab} \, {R}_{mn} - \tilde{R} \, g_{mn} \, {R}_{ab} + \frac{1}{4} \, g_{ab}  \,  g_{mn} \, \tilde{R}^2 \biggr) g .
\end{aligned}
\end{equation}

\noindent Using (\ref{SFD-Linear}), the nondistributional term (\ref{SFD-WavefunctionFuncDerivND1B}) simplifies to:
\begin{equation}\label{SFD-FuncDerivDefn2ND2}
\begin{aligned}
\tilde{G}^{abmn}(y) \left.\text{ND}\biggl\llbracket \frac{\delta^2 \Psi}{\delta g^{ab}_y \,  \delta g^{mn}_z} \biggr\rrbracket\right|_{z=y}
&=\frac{1}{4} \frac{\partial^2 \Psi}{\partial S^2_{\lambda}} \left( 3\, R^2  - 8 \, {R}^{mn} \, {R}_{mn} - 4 \, R \, \lambda+ 12 \, \lambda^2  \right) g .
\end{aligned}
\end{equation}

\noindent Finally, plugging Eqs. (\ref{SFD-FuncDerivInt3c}) and (\ref{SFD-FuncDerivDefn2ND2}) into Eq. (\ref{SFD-WdWHamReg2}), I obtain the following result,
\begin{equation}\label{SFD-WavefunctionFuncDerivFinalExact}
\begin{aligned}
{G}^{abmn} \, \frac{\delta^2 \Psi}{\delta g^{ab}_y \,  \delta g^{mn}_y} & \cong \frac{\kappa^2}{2}  \left(-\frac{1}{2 \, V}\frac{\partial \Psi}{\partial S_{\lambda}} \,  \left( 38 \, R - 15 \, \lambda \right)\sqrt{g} + \frac{\partial^2 \Psi}{\partial S^2_{\lambda}} \left( 3\, R^2  - 8 \, {R}^{mn} \, {R}_{mn} - 4 \, R \, \lambda+ 12 \, \lambda^2  \right) \sqrt{g}\right),
\end{aligned}
\end{equation}

\noindent where [recalling Eq. (\ref{SFD-GMetric2})] I have made use of the relation ${2 \, \kappa^2} \, \tilde{G}^{abmn} ={\sqrt{g}} \, G^{abmn}$.

\subsection{An approximate solution to the Wheeler-DeWitt equation}\label{Sec-SFDApproxSol}
The result (\ref{SFD-WavefunctionFuncDerivFinalExact}) may be used to obtain approximate solutions to the regularized Wheeler-DeWitt equation (\ref{SFD-WdWReg}). First, I set $V=v_0$, where $v_0$ is a constant. For later convenience, I wish to work in terms of a dimensionless parameter; since $v_0$ has units of volume, it is helpful to write $v_0$ in terms of the Planck volume $( \hbar \,  \kappa)^{3/2}$ and a dimensionless parameter $b$:
\begin{equation}\label{SFD-DimensionlessCutoff}
v_0 = b^3 ( \hbar \,  \kappa)^{3/2}.
\end{equation}

\noindent To solve the regularized Wheeler-DeWitt equation, I take a low-curvature limit and neglect terms to second order in the Ricci curvature,\footnote{This is essentially a small curvature expansion for the Wheeler-DeWitt equation, which was introduced in \cite{Hamberetal2013}.} so that Eq. (\ref{SFD-WavefunctionFuncDerivFinalExact}) yields
\begin{equation}\label{SFD-WavefunctionFuncDerivApprox}
\begin{aligned}
{G}^{ijkl} \, \frac{\delta^2 \Psi}{\delta g^{ij}_y \,  \delta g^{kl}_y} 
& \approx \frac{\kappa^2}{2}  \left(-\frac{1}{2 \, b^3 ( \hbar \,  \kappa)^{3/2}}\frac{\partial \Psi}{\partial S_{\lambda}} \,  \left( 38 \, R - 15 \, \lambda \right) - \frac{\partial^2 \Psi}{\partial S^2_{\lambda}} \left(4 \, R \, \lambda - 12 \, \lambda^2  \right)\right)\sqrt{g}\\
\end{aligned}
\end{equation}

\noindent With some algebra, the regularized Wheeler-DeWitt equation (\ref{SFD-WdWReg}) becomes
\begin{equation}\label{SFD-WdWApprox1b}
\begin{aligned}
- \left(2 \, \hbar^2 \, \kappa^2 \, \lambda \, \frac{\partial^2 \Psi}{\partial S^2_{\lambda}} + \frac{19 \, \sqrt{\hbar \, \kappa}}{2 \, b^3}\frac{\partial \Psi}{\partial S_{\lambda}} - \Psi \right) R \, \sqrt{g} +  \frac{1}{2}  \left( \frac{15 \lambda \, \sqrt{\hbar \, \kappa}}{2 \, b^3}\frac{\partial \Psi}{\partial S_{\lambda}} +12 \, \hbar^2 \, \kappa^2 \, \lambda ^2 \, \frac{\partial^2 \Psi}{\partial S^2_{\lambda}} - 4 \, \Lambda \, \Psi \right)\sqrt{g} = 0 .
\end{aligned}
\end{equation}

For simplicity, I first consider the $\lambda=\Lambda=0$ case; here, the Wheeler-DeWitt equation simplifies considerably:
\begin{equation}\label{SFD-WdWApprox1c}
- \left(\frac{19 \, \sqrt{\hbar \, \kappa}}{2 \, b^3}\frac{\partial \Psi}{\partial S_{EH}} - \Psi \right) R \, \sqrt{g} = 0
\end{equation}

\noindent Equation (\ref{SFD-WdWApprox1c}) admits the following solution:
\begin{equation}\label{SFD-WdW2aSolnZero}
\Psi_{0}[g^{\cdot \cdot}]= C_{0} \, \exp\left[\frac{2 \, b^3}{19 \, \sqrt{\hbar \, \kappa}} \, S_{EH}[g^{\cdot \cdot}]\right].
\end{equation}

\noindent Before proceeding to the $\lambda \neq 0$, $\Lambda \neq 0$ case, I argue for the necessity of taking the low-curvature limit, in which I neglect terms quadratic in the Ricci curvature. Recall that the volume average regularization used to obtain Eqs. (\ref{SFD-WavefunctionFuncDerivFinalExact}) and (\ref{SFD-WavefunctionFuncDerivFinalExact2}) is based on the assumption that the Wheeler-DeWitt equation is a low energy description for the effective field theory that results after one has integrated out short distance (large curvature) modes for some theory of quantum gravity. If I reinsert (\ref{SFD-WdW2aSolnZero}) into (\ref{SFD-WavefunctionFuncDerivFinalExact}) and multiply through by $\hbar^3 \, \kappa $, I obtain (setting $\lambda=0$)
\begin{equation}\label{SFD-WavefunctionFuncDerivFinalExact2}
\begin{aligned}
\hbar^3 \, \kappa \, {G}^{abmn} \, \frac{\delta^2 \Psi}{\delta g^{ab}_y \,  \delta g^{mn}_y} & \cong \frac{1}{2}  \biggl(- 2 \,  \hbar \, \kappa \,  R \, \sqrt{g} + \frac{4}{19^2} \, b^6 \, \hbar^2 \, \kappa^2 \, \left( 3\, R^2  - 8 \, {R}^{mn} \, {R}_{mn} \right) \sqrt{g}\biggr)\Psi[g^{\cdot \cdot}].
\end{aligned}
\end{equation}

\noindent If $b$ is on the order of unity (which corresponds to choosing $v_0$ to be on the order of the Planck volume), the limit in which one can neglect the quadratic curvature terms corresponds to the limit $\hbar \, \kappa \, |R_{ab}| \ll 1$, or when the Ricci curvature is much less than the inverse Planck area $(\hbar \, \kappa)^{-1}$. Curvatures on the order of the inverse Planck area correspond to short distance behavior, and it follows that the low-curvature limit is necessary if one chooses the averaging volume $v_0$ to be on the order of the Planck volume $(\hbar \, \kappa)^{3/2}$.

For the $\lambda \neq 0$, $\Lambda \neq 0$ case, I can solve the Wheeler-DeWitt equation (\ref{SFD-WdWApprox1b}) by seeking a function $\Psi(S_{\lambda})$ that satisfies the following set of ordinary differential equations:
\begin{equation}\label{SFD-WdWApprox2a}
2 \, \hbar^2 \, \kappa^2 \, \lambda \, \frac{\partial^2 \Psi}{\partial S^2_{\lambda}} + \frac{19 \, \sqrt{\hbar \, \kappa}}{2 \, b^3}\frac{\partial \Psi}{\partial S_{\lambda}} - \Psi  = 0
\end{equation}
\begin{equation}\label{SFD-WdWApprox2b}
\frac{5 \lambda \, \sqrt{\hbar \, \kappa}}{b^3}\frac{\partial \Psi}{\partial S_{\lambda}} +8 \, \hbar^2 \, \kappa^2 \, \lambda ^2 \, \frac{\partial^2 \Psi}{\partial S^2_{\lambda}} - \frac{8}{3} \, \Lambda \, \Psi = 0 .
\end{equation}

\noindent The first equation (\ref{SFD-WdWApprox2a}) admits solutions of the following form:
\begin{equation}\label{SFD-WdW2aSoln}
\Psi_{\pm}(S_{\lambda}) = C_{\pm} \, \exp\left[\left(\frac{ - 19 \pm \sqrt{361 + 32 \, \lambda \, b^6 \, \hbar \, \kappa}}{8 \,\lambda \, b^3 \, (\hbar \, \kappa)^{3/2}} \right) S_{\lambda}\right]
\end{equation}

\noindent where $C_{+}$ and $C_{-}$ are complex constants. In the $\lambda \rightarrow 0$ limit, $\Psi_{+}$ becomes the $\lambda=0$ solution $\Psi_0$ (\ref{SFD-WdW2aSolnZero}). Inserting $\Psi_{\pm}$ (\ref{SFD-WdW2aSoln}) into Eq. (\ref{SFD-WdWApprox2b}), I obtain the following condition on the parameter $\lambda$:
\begin{equation}\label{SFD-WdW2bSolnlambdaConstraintP}
99 \left(19 - Q \, \sqrt{361+ 32  \, b^6 \, \hbar \, \kappa \, \lambda }\right)  + 32 \, b^6 \, \hbar \, \kappa \left(3 \, \lambda-2 \, \Lambda \right) = 0 .
\end{equation}

\noindent where $Q=+1$ for $\Psi_{+}$ and $Q=-1$ for $\Psi_{-}$. Solving for $\lambda$, I find that for both $Q=1$ and $Q=-1$, I obtain the following values for $\lambda$:
\begin{equation}\label{SFD-WdW2bSolnlambdaP}
\lambda = \frac{128 \, b^6 \, \hbar \, \kappa \, \Lambda -33 \left(15 \pm \sqrt{768 \, b^6 \, \hbar \, \kappa \, \Lambda +225}\right)}{192 \, b^6 \, \hbar \, \kappa }.
\end{equation}

Though Eqs. (\ref{SFD-WdW2aSoln}) and (\ref{SFD-WdW2bSolnlambdaP}) describe a solution to the regularized Wheeler-DeWitt equation, they are unsatisfactory in their present form due to their dependence on the regularization parameter $b$. While one might expect $b \sim 1$, so that $v_0$ is on the order of the Planck volume, the precise value for $b$ is dependent on the details of the short distance physics. On the other hand, the viewpoint here is that quantum general relativity (and by extension quantum geometrodynamics) is a low energy effective field theory, which can be formulated without reference to the details of short distance physics; it is, therefore, appropriate to seek results that are independent of the value for the regularization parameter $b$. 

To obtain a regularization independent result, I recall that the volume averaging regularization was introduced to avoid a delta function divergence, and note that divergences reappear in the Wheeler-DeWitt equation when taking the limit $b \rightarrow 0$, which corresponds to the limit in which the averaging volume goes to zero. I also recall that in perturbative quantum field theory, the coupling constants in the (unrenormalized) action are bare constants that do not correspond to physically meaningful quantities and that in renormalization, one absorbs the divergences into the coupling constants by replacing the bare coupling constants with coupling constants that depend on the regularization parameter (which is effectively what is done with the addition of counterterms in the action). With this in mind, I imagine that $\kappa$ represents a ``bare'' quantity, and introduce a dependence on the regularization parameter $b$. I then require that for small $b$, $\kappa$ has the following leading-order dependence on $b$:
\begin{equation}\label{SFD-RenormalizedKappa}
\kappa=\tilde{\kappa} \, {b^6} + O(b^7)
\end{equation}

\noindent For the $\lambda=\Lambda=0$ solution (\ref{SFD-WdW2aSolnZero}), it is straightforward to see that in the limit $b \rightarrow 0$, Eq. (\ref{SFD-RenormalizedKappa}) for $\kappa$ yields the result:\footnote{Note that the limit $b \rightarrow 0$ provides further justification for dropping the curvature squared terms in (\ref{SFD-WavefunctionFuncDerivFinalExact2}).}
\begin{equation}\label{SFD-WdW2aSolnZeroRenormalized}
\lim_{b \rightarrow 0}\Psi_{0}[g^{\cdot \cdot}]= C_{0} \, \exp\left[\frac{2}{19 \, \sqrt{\hbar \, \tilde{\kappa}}} \, S_{EH}[g^{\cdot \cdot}]\right].
\end{equation}

\noindent For the $\lambda \neq 0$, $\Lambda \neq 0$ solution $\Psi_{+}$ (\ref{SFD-WdW2aSoln}), the limit $b \rightarrow 0$ yields a similar result:\footnote{The exponent becomes infinite in the $\Psi_{-}$ case.}
\begin{equation}\label{SFD-WdWSolnRenormalizedExp}
\lim_{b \rightarrow 0}\Psi_{+}[g^{\cdot \cdot}]  = C_{+} \, \exp\left[\frac{2}{19 \, \sqrt{\hbar \, \tilde{\kappa}}} \, S_{\lambda}[g^{\cdot \cdot}]\right].
\end{equation}

\noindent Taking the same limit for the expression for $\lambda$ in Eq. (\ref{SFD-WdW2bSolnlambdaP}), I find that in the ``$-$'' case, I obtain a finite result that is independent of $\tilde{\kappa}$:
\begin{equation}\label{SFD-LongDistancelambda}
\lambda =\frac{76 \, \Lambda}{15}
\end{equation}

\noindent Thus, in the long-distance limit ($b \rightarrow 0$), $\Psi_{+}$ has the explicit form:
\begin{equation}\label{SFD-WdWSolnRenormalized}
\Psi[g^{\cdot \cdot}] = A \, \exp\left[\frac{2}{19 \, \sqrt{\hbar \, \tilde{\kappa}}}\left(S_{EH}[g^{\cdot \cdot}] - \frac{152}{15} \, \Lambda \, \mathcal{V}[g^{\cdot \cdot}]\right)\right] .
\end{equation}

\noindent Eq. (\ref{SFD-WdWSolnRenormalized}) forms the main result of this article; it describes a solution of the regularized Wheeler-DeWitt equation in the low-curvature, long-distance limit.

One may note that the solution (\ref{SFD-WdWSolnRenormalized}) has a form similar to the that proposed in Eq. (107) of \cite{Hamberetal2013} for the large-volume limit. Unfortunately, the solution (\ref{SFD-WdWSolnRenormalized}), obtained from a Regge simplicial lattice regularization, is distinct from the large-volume solution presented in Eqs. (107), (117), and (118) of \cite{Hamberetal2013}, so a direct comparison cannot be made. In particular, the coefficients (Eq. (118) of \cite{Hamberetal2013}) in front of the volume functional and the Einstein-Hilbert functional in their solution differ by a factor of $i$, and have a different dependence on the value of $\Lambda$; in the large-volume the solution presented in \cite{Hamberetal2013}, the coefficient for the volume functional vanishes in the limit $\Lambda \rightarrow 0$, the coefficient in front of the Einstein-Hilbert functional diverges.\footnote{This remark is not meant to be a criticism; I am merely pointing out the differences between the solution presented in this article and the solution presented in \cite{Hamberetal2013} that preclude a direct comparison.}

\subsection{A three-sphere universe}
I conclude this article with a brief investigation of the solution described in Eq. (\ref{SFD-WdWSolnRenormalized}) for a minisuperspace restriction to the ``round'' geometry for a three-sphere, given by the line element
\begin{equation}\label{EHWF-3SphereMetric}
ds^2
=r^2 \left[d\psi^2 +\sin^2 \, \psi \> \left( d\theta^2 + \sin^2 \, \theta \> d\phi^2\right) \right]
\end{equation}

\noindent For the round metric (\ref{EHWF-3SphereMetric}) on the three-sphere, $ S_{EH}=12 \, \pi^2 \, r$, and $\mathcal{V}=2 \, \pi^2 \, r^3$. The wavefunctional (\ref{SFD-WdWSolnRenormalized}) evaluated for the geometry (\ref{EHWF-3SphereMetric}) is given by
\begin{equation}\label{SFD-WdWSoln3Sphere}
\Psi(r) = A \, \exp\left[ \frac{24 \, \pi^2}{19 \, \sqrt{\hbar \, \tilde{\kappa}}} \left(r - \frac{76}{45} \, \Lambda \, r^3\right)\right] .
\end{equation}

\noindent I note that for $\Lambda > 0$, $\lim_{r\rightarrow \infty} \Psi(r) = 0$, and that $\Psi(r)$ has a maximum at $r = \sqrt{15/76 \, \Lambda}$; in a DeSitter universe with a closed slicing, this three-geometry corresponds to a time $t$ satisfying $r^2=3 \cosh^2(t \sqrt{3/\Lambda})/\Lambda$. In the minisuperspace restriction, one can normalize $\Psi(r)$, as long as $\tilde{\kappa}$ and $\Lambda$ have finite values; the integral (performed with Mathematica \cite{Mathematica}) of the square of $\Psi$ has the following form:
\begin{equation}\label{SFD-WdWSoln3SphereInt}
|A|^2 \int_0^\infty \exp \left(\mu \, r - \frac{\nu}{3} \, r^3\right) dr = \frac{|A|^2}{6 \, \nu} \left( 4  \pi  \, \nu^{2/3} \, \text{Bi}({\mu}/{\nu^{1/3}}) +3 \, \mu^2 \, {_1}F{_2}(1;4/3,5/3; \mu^3 /9 \, \nu)\right)
\end{equation}

\noindent where $\text{Bi}(x)$ is an Airy function of the second kind, and ${_p}F{_q}(r_1,...r_p; s_1,...s_q ;x)$ is a generalized hypergeometric function. Though the result diverges for $\nu \rightarrow 0$ (which corresponds to taking $\Lambda \rightarrow 0$), the above remains finite for finite values of the parameters $\mu$ and $\nu$. The divergence in the limit $\nu \rightarrow 0$ corresponds to setting $\Lambda=0$; in this case, one can see that  $\lim_{r\rightarrow \infty} \Psi(r) = \infty$. This suggests that for the three-sphere manifold, the state $\Psi[g^{\cdot \cdot}]$ (\ref{SFD-WdWSolnRenormalized}) is not normalizable for $\Lambda=0$. One might observe that the unboundedness for $\Psi(r)$ when $\Lambda=0$ corresponds to the limit in which the volume becomes infinite. In minisuperspace models, the scale factor of the FRW metric, which controls the volume for spatial slices, often plays the role of a time parameter \cite{Misner1969,Kiefer2009,Kiefer2012QG}. One may attempt to resolve the unboundedness in the (nonminisuperspace) functional $\Psi[g^{\cdot \cdot}]$ by treating the volume $V_{\Sigma}$ of the three-manifold $\Sigma$ as a time parameter; however, while this might lead to a normalizable state at a fixed volume, it does so at the cost of nonunitary time evolution \cite{GambiniPorto2001,*GrybThebault2018}.


\section{Final Remarks}
In this article, I have examined a volume average regularization for the second functional derivative operator in the Wheeler-DeWitt equation. I have argued that such a regularization is natural for studying quantum geometrodynamics if one regards quantum general relativity to be a low energy effective field theory of quantum gravity. In the low-curvature, long-distance limit, I have found a solution [Eq. (\ref{SFD-WdWSolnRenormalized})] to the regularized Wheeler-DeWitt equation.

An important question is whether the solution $\Psi[g^{\cdot \cdot}]$ Eq. (\ref{SFD-WdWSolnRenormalized}) describes a physically meaningful state for quantum geometrodynamics. I have briefly studied the features of $\Psi[g^{\cdot \cdot}]$ Eq. (\ref{SFD-WdWSolnRenormalized}) for three-sphere geometries, and have found that for finite $\tilde{\kappa}$ and finite $\Lambda>0$, the solution is normalizable on the minisuperspace restriction to metrics of the form (\ref{SFD-WdWSoln3Sphere}). It is curious that the minisuperspace state $\Psi(r)$ (\ref{SFD-WdWSoln3Sphere}) is peaked at the geometry corresponding to a particular time in the closed slicing of DeSitter spacetime; this seems to suggest that the state described by $\Psi[g^{\cdot \cdot}]$ Eq. (\ref{SFD-WdWSolnRenormalized}) contains some information about the temporal placement of the three-geometry in spacetime which in turn suggests that a more complete account of the dynamics requires solutions with a more complicated dependence on the three-geometry. One difficulty, as discussed earlier is the unboundedness of $\Psi[g^{\cdot \cdot}]$ for geometries on the three-sphere manifold when $\Lambda=0$. One can, however, place an upper bound on $S_{EH}[g^{\cdot \cdot}]$ with certain choices of topology on compact manifolds; in fact, it has been shown \cite{YauSchoen1979} that $S_{EH}[g^{\cdot \cdot}]$ always has a negative value for the three-torus.\footnote{An interesting question is whether one can identify other three-manifolds that have this property---in particular, one seeks three-manifolds with a negative or vanishing Yamabe (topological) invariant \cite{LeeParker1987}, which implies $S_{EH}[g^{\cdot \cdot}] \leq 0$.} 

There are some general issues that have not been addressed in this article, some which have been discussed elsewhere in the literature, and some which I leave for future work. A particularly intriguing line of investigation, which I leave for future work, concerns the $V=V_{\Sigma}$ quantum gravity model briefly described in Sec. \ref{Sec-SFDOPWDW}. Another question of interest is whether it is appropriate to replace the second functional derivative operator in the Wheeler-DeWitt equation (\ref{SFD-WdW}) with a Laplace-Beltrami operator, such as those described in \cite{DeWitt1967} and \cite{FengMatzner2017prd}. Though the methods presented in this article are motivated by effective field theory considerations and inspired by renormalization theory, the precise relationship between the methods presented here and perturbative quantum field theory is presently unclear. In particular, the methods presented here are formally nonperturbative and gauge/slicing dependent\footnote{of particular concern is the fact that low 3-curvature limit used to obtain the approximate solutions is gauge/slicing dependent}, which complicate the task of establishing the relationship between the results presented in this article and relativistic quantum field theory. One difficulty in particular concerns the fact that the volume averaging is performed over a spatial volume, rather than a spacetime volume; to fully establish the relationship between the regularization presented in this article to a covariant regularization, one may be required to perform an additional temporal averaging, in which one must confront the problem of time. Furthermore, one must take into consideration the fact that $\Psi_\alpha[g^{\cdot \cdot}]$ are approximate\footnote{Though as argued in Sec. \ref{Sec-SFDApproxSol}, such an approximation is necessary if one considers the reasoning used to justify the volume average regularization.}
solutions to an equation that is only valid in a low-energy limit---in the effective field theory framework, the Wheeler-DeWitt equation itself is only valid at scales in which quantum general relativity remains valid; in particular, the solution is only expected to be valid at scales where one can ignore the effects of curvature-squared terms in the bulk (four-dimensional) action. Since the approximate solutions $\Psi_\alpha[g^{\cdot \cdot}]$ are functionals of $g^{ij}$, they automatically contain information at all scales \cite{Carlip2017}. This may require the suppression of information contained in $\Psi_\alpha[g^{\cdot \cdot}]$ for 3-geometries corresponding to scales where the Wheeler-DeWitt equation is no longer expected to be valid. 

\begin{acknowledgments}
This work was partially supported by the National Science Foundation under Grant No. PHY-1620610. I thank Richard Matzner for his questions and remarks, which helped to sharpen both my thoughts and my writing. I also thank Mark Selover, Philip Morrison, and Baruch Garcia for their comments and feedback.
\end{acknowledgments}

\noindent\textit{Note added in proof.}---I recently became aware of another set of approaches in the literature which regularize the Wheeler-DeWitt equation. I am referring in particular to the heat kernel and point splitting regularizations described in \cite{Horiguchietal1995,MaedaetalPRD1996,Kowalskietal1996,Mansfield1994,WaldPRD1978}, which are related to those described in the present article; the volume average regularization may be intepreted as an average over the displacement in the point splitting regularization.

\appendix*


\section{The variation of the Einstein-Hilbert action} \label{AppendixVariationCurvatureEH}

\subsection{The change in the ricci tensor}
In this section, I review the variation of the Ricci tensor. In particular, I work out the change in the curvature tensor under the following transformation of the connection coefficients,
\begin{equation} \label{A-ConnectionTransformed}
\tilde{\Gamma}^a_{ij}=\Gamma^a_{ij}+Q^a_{ij} \>\>\>\>\> \Rightarrow  \>\>\>\>\>  Q^a_{ij} = \Delta {\Gamma}^a_{ij},
\end{equation}

\noindent where $Q^a_{ij}$ are components of a tensor. The transformed Riemann curvature tensor may be written
\begin{equation}\label{A-CurvatureTensorTransformedAdditiveA}
{\tilde{R}^i}{_{jab}}={R^i}{_{jab}}+\partial_a Q^i_{bj}-\partial_b Q^i_{aj}+Q^i_{as} \>\Gamma^s_{bj}+\Gamma^i_{as} \>Q^s_{bj}- Q^i_{bs} \>\Gamma^s_{aj}- \Gamma^i_{bs} \>Q^s_{aj}+Q^i_{as} \>Q^s_{bj}- Q^i_{bs} \>Q^s_{aj} .
\end{equation}

\noindent Comparing this with the covariant derivatives of $Q^a_{ij}$,
\begin{equation}\label{A-CovariantDerivativeQ}
\begin{aligned}
\nabla_a Q^i_{bj}&=\partial_a Q^i_{bj}+\Gamma^i_{as}Q^s_{bj}-\Gamma^s_{ab}Q^i_{sj}-\Gamma^s_{aj}Q^i_{bs}\\
\nabla_b Q^i_{aj}&=\partial_b Q^i_{aj}+\Gamma^i_{bs}Q^s_{aj}-\Gamma^s_{ba}Q^i_{sj}-\Gamma^s_{bj}Q^i_{as},
\end{aligned}
\end{equation}

\noindent I find that
\begin{equation}\label{A-CurvatureTensorTransformedAdditive}
{\tilde{R}^i}{_{jab}}-{R^i}{_{jab}}+S^s_{ab} \>Q^i_{sj}=\nabla_a Q^i_{bj}-\nabla_b Q^i_{aj}+Q^i_{as} \>Q^s_{bj}- Q^i_{bs} \>Q^s_{aj},
\end{equation}

\noindent where $S^s_{ab}:=\Gamma^s_{ab}-\Gamma^s_{ba}$ is the torsion tensor. The torsion tensor comes from the fact that the terms $\Gamma^s_{ab}Q^i_{sj}$ and $\Gamma^s_{ba}Q^i_{sj}$ in the covariant derivatives (\ref{A-CovariantDerivativeQ}) do not appear in the expression (\ref{A-CurvatureTensorTransformedAdditiveA}) and must be added in when converting the partial derivatives of the connection variations to covariant derivatives. I contract indices to also obtain the transformation of the Ricci tensor:
\begin{equation}\label{A-RicciTensorTransformedAdditiveQ}
{\tilde{R}}_{ab}-{R}_{ab}+S^s_{ib} \>Q^i_{sa}=\nabla_i Q^i_{ba}-\nabla_b Q^i_{ia}+Q^i_{is} \>Q^s_{ba}- Q^i_{bs} \>Q^s_{ia}.
\end{equation}

\noindent Recalling $Q^a_{ij} = \Delta {\Gamma}^a_{ij}$  (\ref{A-ConnectionTransformed}), I may rewrite the above as
\begin{equation}\label{A-RicciTensorTransformedAdditive}
{\tilde{R}}_{ab}-{R}_{ab}+S^s_{ib} \>\Delta {\Gamma}^i_{sa}=\nabla_i \Delta {\Gamma}^i_{ba}-\nabla_b \Delta {\Gamma}^i_{ia}+\Delta {\Gamma}^i_{is} \>\Delta {\Gamma}^s_{ba}- \Delta {\Gamma}^i_{bs} \>\Delta {\Gamma}^s_{ia}.
\end{equation}

\noindent For a torsion-free connection, $S^s_{ib}=0$, I may rewrite (\ref{A-RicciTensorTransformedAdditive}) as
\begin{equation}\label{A-RicciTensorVariation}
\begin{aligned}
\Delta {R}_{ab} &:={\tilde{R}}_{ab}-{R}_{ab}  = \nabla_i \Delta {\Gamma}^i_{ba}-\nabla_b \Delta {\Gamma}^i_{ia}+\Delta {\Gamma}^i_{is} \>\Delta {\Gamma}^s_{ba}- \Delta {\Gamma}^i_{bs} \>\Delta {\Gamma}^s_{ia}.
\end{aligned}
\end{equation}

\subsection{The variation of the Einstein-Hilbert action to first order}
I now review the first-order variation of the Einstein-Hilbert action, which may be found in a standard text on general relativity \cite{MTW,Wald,Carroll}. The variation of the Ricci tensor is
\begin{equation}\label{SFD-CurvatureTensorVariation0}
\delta {R}_{ab} = \nabla_i \delta \Gamma^i_{ba}-\nabla_b \delta \Gamma^i_{ia},
\end{equation}

\noindent where
\begin{equation} \label{SFD-ChristoffelSymbolVariationTerms}
\begin{aligned}
\delta \Gamma^i_{ia}&=\frac{1}{2} \> g^{is} (\nabla_i \delta g_{sa}+\nabla_a \delta g_{is}-\nabla_s \delta g_{ia})\\
\delta \Gamma^i_{ba}&=\frac{1}{2} \> g^{is}(\nabla_b \delta g_{sa}+\nabla_a \delta g_{bs}-\nabla_s \delta g_{ba}).
\end{aligned}
\end{equation}

\noindent The variation of the Ricci tensor takes the following form,
\begin{equation}\label{SFD-CurvatureTensorVariation1}
2\delta {R}_{ab} = g^{ij} \left(  \nabla_i \nabla_b \delta g_{ja} +  \nabla_i \nabla_a \delta g_{bj} -  \nabla_i \nabla_j \delta g_{ba} - \nabla_b \nabla_a \delta g_{ij}\right),
\end{equation}

\noindent and it follows that the variation of the Ricci scalar is
\begin{equation}\label{SFD-RicciScalarVariationB}
\begin{aligned}
\delta {R}&:= g^{ab}\>\delta {R}_{ab} + {R}_{ab} \> \delta g^{ab} = g^{ij} \> g^{ab} \> \nabla_i (\nabla_b \delta g_{ja} - \nabla_j \delta g_{ab})+{R}_{ab}\> \delta g^{ab}.
\end{aligned}
\end{equation}

\noindent To first order, the variation of the metric and its inverse are related in the following manner:
\begin{equation} \label{SFD-InverseConditionVariationA}
\begin{aligned}
\delta g_{ab} = - g_{a m} \> g_{bn} \> \delta g^{mn}.
\end{aligned}
\end{equation}

\noindent I use the above (\ref{SFD-InverseConditionVariationA}) to rewrite equations (\ref{SFD-CurvatureTensorVariation1}) and (\ref{SFD-RicciScalarVariationB}) as
\begin{equation}\label{SFD-CurvatureTensorVariation2}
2\delta {R}_{ab} = - g^{ij} \left(  g_{j m} \> g_{an} \nabla_i \nabla_b \delta g^{mn} +   g_{b m} \> g_{jn} \nabla_i \nabla_a \delta g^{mn} -  g_{a m} \> g_{bn}  \nabla_i \nabla_j \delta g^{mn} -  g_{i m} \> g_{jn} \nabla_b \nabla_a \delta g^{mn}\right),
\end{equation}
\begin{equation}\label{SFD-CurvatureTensorVariation2b}
\delta {R}= - (\nabla_i \nabla_j \delta g^{ij} - g_{mn} \, g^{ij} \, \nabla_i \nabla_j \delta g^{mn} )+{R}_{ab}\> \delta g^{ab}.
\end{equation}

\noindent Using (\ref{SFD-VolumeElementVariationASO}), the variation of the volume element is, to first order,
\begin{equation} \label{SFD-VolumeElementVariationA}
\begin{aligned}
\delta \sqrt{g}= \frac{1}{2} \sqrt{g} \> g^{nm} \> \delta g_{mn}=-\frac{1}{2} \sqrt{g} \> g_{mn} \> \delta g^{mn}.
\end{aligned}
\end{equation}

\noindent I now present the algebra for the first variation of the Einstein-Hilbert action,
\begin{equation}\label{A-3EH1Var}
\begin{aligned}
\delta S_{EH}&=\int_{\Sigma} \, \left( \delta R \, \sqrt{g} +R \, \delta\sqrt{g} \right) d^3 y\\
&=\int_{\Sigma} \, \left( \left( - g^{ij} \> g^{ab} \> \nabla_i (g_{j m} \> g_{an} \nabla_b \delta g^{mn} - g_{a m} \> g_{bn} \nabla_j \delta g^{mn})+{R}_{ab}\> \delta g^{ab} \right)\, \sqrt{g} - \frac{1}{2} \sqrt{g} \> g_{ab} \> \delta g^{ab} \, R \right) d^3 y\\
&=\int_{\Sigma} \, \left({R}_{ab} - \frac{1}{2}  \> g_{ab} \> R \right) \delta g^{ab} \, \sqrt{g} \,  d^3 y,
\end{aligned}
\end{equation}

\noindent where a boundary term has been dropped in the second equality due to the fact that the manifold $\Sigma$ is compact and without boundary (recall the metric compatibility condition $\nabla_k g_{ij}=0$,  $\nabla_k g^{ij}=0$).

\subsection{The variation of the Einstein-Hilbert action to second order}
Here, I present some algebra for the variation of the Einstein-Hilbert action leading up to Eq. (\ref{SFD-3EHVar2}). First, I expand the variation of the Einstein-Hilbert action,
\begin{equation}\label{A-3EHVar2O}
\begin{aligned}
\Delta S_{EH}  &:= S_{EH}[g^{\cdot \cdot} + \delta g^{\cdot \cdot}] - S_{EH}[g^{\cdot \cdot}]\\
&= \int_{\Sigma} \left( \Delta R \, \sqrt{g} + R \, \Delta \sqrt{g} + \Delta R \, \Delta \sqrt{g}\right) d^3 y\\
&= \int_{\Sigma} \biggl( \left[  \delta g^{ab} \,  {R}_{ab} + g^{ab} \, \Delta {R}_{ab} + \delta g^{ab} \, \Delta {R}_{ab} \right] - \frac{1}{2} R \, \left[g_{ab} \>  \delta g^{a b} - Y_{abmn} \, \delta g^{ab} \, \delta g^{mn} \right]  -  \frac{1}{2}  \,g_{mn} \, \Delta R \, \delta g^{mn} \\
&\>\>\>\>\>\>\>\>\>\>\>\>\>\>\> + \mathcal{O}([\delta g^{\cdot \cdot}]^3)\biggr) \sqrt{g} \> d^3 y\\
&= \int_{\Sigma} \biggl( \left[  \delta g^{ab} \,  {R}_{ab} + g^{ab} \, \Delta {R}_{ab} + \delta g^{ab} \, \Delta {R}_{ab} \right]  - \frac{1}{2} R \, g_{ab} \>  \delta g^{a b} +\frac{1}{2} \, R  \, Y_{abmn} \, \delta g^{ab} \, \delta g^{mn} \\
&\>\>\>\>\>\>\>\>\>\>\>\>\>\>\> -  \frac{1}{2}  \,g_{mn} \, \delta g^{mn} \{  \delta g^{ab} \,  {R}_{ab} + g^{ab} \, \Delta {R}_{ab} \} + \mathcal{O}([\delta g^{\cdot \cdot}]^3)\biggr) \sqrt{g} \> d^3 y .
\end{aligned}
\end{equation}

\noindent Next, I substitute the expression for $\Delta {R}_{ab}$ in Eq. (\ref{A-RicciTensorVariation}) into Eq. (\ref{A-3EHVar2O}) to obtain (keeping terms to second order in variations):
\begin{equation}\label{A-3EHVar2a}
\begin{aligned}
\Delta S_{EH} &=\int_{\Sigma} \biggl( \left\{  \delta g^{ab} \,  {R}_{ab} + (g^{ab} + \delta g^{ab}) \, \left[\nabla_i \Delta {\Gamma}^i_{ba}-\nabla_b \Delta {\Gamma}^i_{ia}+\Delta {\Gamma}^i_{is} \>\Delta {\Gamma}^s_{ba}- \Delta {\Gamma}^i_{bs} \>\Delta {\Gamma}^s_{ia}\right] \right\}  - \frac{1}{2} R \, g_{ab} \>  \delta g^{a b}\\
&\>\>\>\>\>\>\>\>\>\>\>\>\>\>\> +\frac{1}{2} \, R  \, Y_{abmn} \, \delta g^{ab} \, \delta g^{mn}  -  \frac{1}{2}  \,g_{mn} \, \delta g^{mn} \left\{  \delta g^{ab} \,  {R}_{ab} + g^{ab} \, \left[\nabla_i \Delta {\Gamma}^i_{ba}-\nabla_b \Delta {\Gamma}^i_{ia}+\Delta {\Gamma}^i_{is} \>\Delta {\Gamma}^s_{ba}- \Delta {\Gamma}^i_{bs} \>\Delta {\Gamma}^s_{ia}\right] \right\}\\
&\>\>\>\>\>\>\>\>\>\>\>\>\>\>\> + \mathcal{O}([\delta g^{\cdot \cdot}]^3)\biggr) \sqrt{g} \> d^3 y\\
&=\int_{\Sigma} \biggl( \left\{  \delta g^{ab} \,  {R}_{ab} + g^{ab}\left[\nabla_i \Delta {\Gamma}^i_{ba}-\nabla_b \Delta {\Gamma}^i_{ia}+\Delta {\Gamma}^i_{is} \>\Delta {\Gamma}^s_{ba}- \Delta {\Gamma}^i_{bs} \>\Delta {\Gamma}^s_{ia}\right] + \delta g^{ab} \left(\nabla_i \Delta {\Gamma}^i_{ba}-\nabla_b \Delta {\Gamma}^i_{ia}\right) \right\} \\
&\>\>\>\>\>\>\>\>\>\>\>\>\>\>\>  - \frac{1}{2} R \, g_{ab} \>  \delta g^{a b} +\frac{1}{2} \, R  \, Y_{abmn} \, \delta g^{ab} \, \delta g^{mn}   -  \frac{1}{2}  \,g_{mn} \, \delta g^{mn}  \delta g^{ab} \,  {R}_{ab} -  \frac{1}{2}  \,g_{mn} \, \delta g^{mn}  g^{ab} \, \left(\nabla_i \Delta {\Gamma}^i_{ba}-\nabla_b \Delta {\Gamma}^i_{ia}\right)\\
&\>\>\>\>\>\>\>\>\>\>\>\>\>\>\> + \mathcal{O}([\delta g^{\cdot \cdot}]^3)\biggr) \sqrt{g} \> d^3 y .
\end{aligned}
\end{equation}

\noindent A rearrangement of terms yields the result [Eq. (\ref{SFD-3EHVar2})]:
\begin{equation}\label{A-3EHVar2b}
\begin{aligned}
\Delta S_{EH} &= \int_{\Sigma} \biggl[ \left( {R}_{ab} - \frac{1}{2} R \, g_{ab} \right)  \delta g^{a b} + g^{ab} \, (\nabla_i \Delta {\Gamma}^i_{ba}-\nabla_b \Delta {\Gamma}^i_{ia} ) + g^{ab} \, \Delta {\Gamma}^i_{is} \>\Delta {\Gamma}^s_{ba} - g^{ab} \, \Delta {\Gamma}^i_{bs} \>\Delta {\Gamma}^s_{ia}\\
&\>\>\>\>\>\>\>\>\>\>\>\>\>\>\>  + \delta g^{ab} \, (\nabla_i \Delta {\Gamma}^i_{ba}-\nabla_b \Delta {\Gamma}^i_{ia}) + \frac{1}{2} \, R \, Y_{abmn} \, \delta g^{ab} \> \delta g^{mn}  -  \frac{1}{2} \,  {R}_{ab} \,g_{mn} \, \delta g^{mn} \, \delta g^{ab}\\
&\>\>\>\>\>\>\>\>\>\>\>\>\>\>\>   -  \frac{1}{2} \, g^{ab} \, g_{mn} \, \delta g^{mn}(\nabla_i \Delta {\Gamma}^i_{ba}-\nabla_b \Delta {\Gamma}^i_{ia}) + \mathcal{O}([\delta g^{\cdot \cdot}]^3)\biggr] \sqrt{g} \> d^3 y.
\end{aligned}
\end{equation}

\noindent Using Eq. (\ref{A-3EH1Var}), I may further simplify this to obtain the result
\begin{equation}\label{A-3EHVar2c}
\begin{aligned}
\Delta S_{EH} = \delta S_{EH} + \int_{\Sigma} \biggl[ &g^{ab} \, (\nabla_i \Delta {\Gamma}^i_{ba}-\nabla_b \Delta {\Gamma}^i_{ia} ) + \frac{1}{2}\left( R \, Y_{abmn} - {R}_{ab} \,g_{mn} \right) \delta g^{ab} \, \delta g^{mn} + g^{ab} \left( \Delta {\Gamma}^i_{is} \>\Delta {\Gamma}^s_{ba} - \Delta {\Gamma}^i_{bs} \>\Delta {\Gamma}^s_{ia} \right)\\
& + \left(\delta g^{ab} -  \frac{1}{2} \, g^{ab} \, g_{mn} \, \delta g^{mn} \right)(\nabla_i \Delta {\Gamma}^i_{ba}-\nabla_b \Delta {\Gamma}^i_{ia})    + \mathcal{O}([\delta g^{\cdot \cdot}]^3)\biggr] \sqrt{g} \, d^3 y.
\end{aligned}
\end{equation}

\subsection{Simplifying terms in the second-order variation of the Einstein-Hilbert action}
In this section, I present the algebra for obtaining Eq. (\ref{SFD-3EHVar2C}) from Eq. (\ref{SFD-3EHVar4}). First, I rewrite Eq. (\ref{SFD-3EHVar4}):
\begin{equation}\label{A-3EHVar4}
\begin{aligned}
\Delta S_{EH} = \delta S_{EH} + \int_{\Sigma} \biggl[& g^{ab} \, \delta {\Gamma}^i_{is} \>\delta {\Gamma}^s_{ba} - g^{ab} \, \delta {\Gamma}^i_{bs} \>\delta {\Gamma}^s_{ia} + \left(\delta g^{ab}  -  \frac{1}{2} \, g^{ab} \, g_{mn} \, \delta g^{mn} \right) (\nabla_i \delta {\Gamma}^i_{ba}-\nabla_b \delta {\Gamma}^i_{ia}) \\
&  + \frac{1}{2} \left( R \, Y_{abmn}  -   {R}_{ab} \,g_{mn}\right) \delta g^{ab} \> \delta g^{mn} \biggr] \sqrt{g} \> d^3 y .
\end{aligned}
\end{equation}

\noindent I apply the divergence theorem to obtain
\begin{equation}\label{A-3EHVar4A}
\begin{aligned}
\Delta S_{EH} & = \delta S_{EH} + \int_{\Sigma} \biggl[ g^{ab} \, \delta {\Gamma}^i_{is} \>\delta {\Gamma}^s_{ba} - g^{ab} \, \delta {\Gamma}^i_{bs} \>\delta {\Gamma}^s_{ia} - \left(\nabla_i \delta g^{ab}  -  \frac{1}{2} \, g^{ab} \, g_{mn} \, \nabla_i \delta g^{mn} \right) \delta {\Gamma}^i_{ba} \\
& \>\>\>\>\>\>\>\>\>\>\>\>\>\>\>\>\>\>\>\>\>\>\>\>\>\>\>\>\>\>  + \left(\nabla_b \delta g^{ab}  -  \frac{1}{2} \, g^{ab} \, g_{mn} \, \nabla_b \delta g^{mn} \right) \delta {\Gamma}^i_{ia} + \frac{1}{2} \left( R \, Y_{abmn}  -   {R}_{ab} \,g_{mn}\right) \delta g^{ab} \> \delta g^{mn} \biggr] \sqrt{g} \> d^3 y \\
& = \delta S_{EH} + \int_{\Sigma} \biggl[ g^{ab} \, \delta {\Gamma}^i_{is} \>\delta {\Gamma}^s_{ba} - g^{ab} \, \delta {\Gamma}^i_{bs} \>\delta {\Gamma}^s_{ia} - \nabla_i \delta g^{ab} \, \delta {\Gamma}^i_{ba}  + \frac{1}{2} \, g^{ab} \, g_{mn} \, \nabla_i \delta g^{mn} \, \delta {\Gamma}^i_{ba} + \nabla_b \delta g^{ab} \, \delta {\Gamma}^i_{ia} \\
& \>\>\>\>\>\>\>\>\>\>\>\>\>\>\>\>\>\>\>\>\>\>\>\>\>\>\>\>\>\>    -  \frac{1}{2} \, g^{ab} \, g_{mn} \, \nabla_b \delta g^{mn}\, \delta {\Gamma}^i_{ia} + \frac{1}{2} \left( R \, Y_{abmn}  -   {R}_{ab} \,g_{mn}\right) \delta g^{ab} \> \delta g^{mn}  \biggr] \sqrt{g} \> d^3 y .
\end{aligned}
\end{equation}

Now the first-order variation of the Christoffel symbols (\ref{SFD-ChristoffelSymbolVariationB}) may be used to obtain the following expressions, which will be useful for working out expressions for (\ref{A-ConnectionScalars}):
\begin{equation} \label{SFD-ChristoffelSymbolVariationC1}
\begin{aligned}
\delta \Gamma^s_{ia} &= \frac{1}{2}(g_{mi} \, g_{na} \, g^{sk}\,  \nabla_k \delta g^{mn} - g_{na} \, \nabla_i \delta g^{sn} - g_{mi} \, \nabla_a \delta g^{ms})\\
\delta \Gamma^s_{ba} &= \frac{1}{2}(g_{mb} \, g_{na} \, g^{sk}\,  \nabla_k \delta g^{mn} - g_{na} \, \nabla_b \delta g^{sn} - g_{mb} \, \nabla_a \delta g^{ms})\\
\delta \Gamma^i_{bs} &= \frac{1}{2}(g_{mb} \, g_{ns} \, g^{ik}\,  \nabla_k \delta g^{mn} - g_{ns} \, \nabla_b \delta g^{in} - g_{mb} \, \nabla_s \delta g^{mi})\\
\delta \Gamma^i_{ia} &= - \frac{1}{2}(g_{mi} \, \nabla_a \delta g^{mi}).
\end{aligned}
\end{equation}

\noindent I use the last one ($\delta \Gamma^i_{ia} = - \frac{1}{2} \, g_{ij} \, \nabla_a \delta g^{ij}$) to simplify some terms in (\ref{A-3EHVar4A}):
\begin{equation}\label{A-3EHVar4A2}
\begin{aligned}
\Delta S_{EH} & = \delta S_{EH} + \int_{\Sigma} \biggl[ - \frac{1}{2} \, g^{ab} \, \, g_{ij} \, \nabla_s \delta g^{ij} \, \delta {\Gamma}^s_{ba} - g^{ab} \, \delta {\Gamma}^i_{bs} \>\delta {\Gamma}^s_{ia} - \nabla_i \delta g^{ab} \, \delta {\Gamma}^i_{ba}  + \frac{1}{2} \, g^{ab} \, g_{mn} \, \nabla_i \delta g^{mn} \, \delta {\Gamma}^i_{ba} \\
& \>\>\>\>\>\>\>\>\>\>\>\>\>\>\>\>\>\>\>\>  - \frac{1}{2} \, g_{ij} \, \nabla_b \delta g^{ab} \, \nabla_a \delta g^{ij} + \frac{1}{4} \, g^{ab} \, g_{mn} \, g_{ij} \, \nabla_b \delta g^{mn}\, \nabla_a \delta g^{ij} + \frac{1}{2} \left( R \, Y_{abmn}  -   {R}_{ab} \,g_{mn}\right) \delta g^{ab} \> \delta g^{mn} \biggr] \sqrt{g} \> d^3 y .
\end{aligned}
\end{equation}

\noindent Now I insert some Kronecker deltas and change index labels so that I can combine terms:
\begin{equation}\label{A-3EHVar4A3}
\begin{aligned}
\Delta S_{EH} 
& = \delta S_{EH} + \int_{\Sigma} \biggl[ - \frac{1}{2} \, g^{ab} \, \, g_{mn} \, \nabla_s \delta g^{mn} \, \delta {\Gamma}^s_{ba} - g^{ab} \, \delta {\Gamma}^i_{bs} \>\delta {\Gamma}^s_{ia} - \nabla_s \delta g^{ab} \, \delta {\Gamma}^s_{ba}  + \frac{1}{2} \, g^{ab} \, g_{mn} \, \nabla_s \delta g^{mn} \, \delta {\Gamma}^s_{ba} \\
& \>\>\>\>\>\>\>\>\>\>\>\>\>\>\>\>\>\>\>\>  - \frac{1}{2} \, g_{ij} \, \nabla_b \delta g^{ab} \, \nabla_a \delta g^{ij} + \frac{1}{4} \, g^{ab} \, g_{mn} \, g_{ij} \, \nabla_b \delta g^{mn}\, \nabla_a \delta g^{ij} + \frac{1}{2} \left( R \, Y_{abmn}  -   {R}_{ab} \,g_{mn}\right) \delta g^{ab} \> \delta g^{mn} \biggr] \sqrt{g} \> d^3 y \\
& = \delta S_{EH} + \int_{\Sigma} \biggl[ \left(\frac{1}{2} \, g^{ab} \, g_{mn} \, \nabla_s \delta g^{mn} - \frac{1}{2} \, g^{ab} \, \, g_{mn} \, \nabla_s \delta g^{mn} - \nabla_s \delta g^{ab} \right) \delta {\Gamma}^s_{ba} - g^{ab} \, \delta {\Gamma}^i_{bs} \>\delta {\Gamma}^s_{ia}  \\
& \>\>\>\>\>\>\>\>\>\>\>\>\>\>\>\>\>\>\>\>  - \frac{1}{2} \, g_{ij} \, \nabla_b \delta g^{ab} \, \nabla_a \delta g^{ij} + \frac{1}{4} \, g^{ab} \, g_{mn} \, g_{ij} \, \nabla_b \delta g^{mn}\, \nabla_a \delta g^{ij} + \frac{1}{2} \left( R \, Y_{abmn}  -   {R}_{ab} \,g_{mn}\right) \delta g^{ab} \> \delta g^{mn} \biggr] \sqrt{g} \> d^3 y \\
& = \delta S_{EH} + \int_{\Sigma} \biggl[  - \nabla_s \delta g^{ab} \, \delta {\Gamma}^s_{ba} - g^{ab} \, \delta {\Gamma}^i_{bs} \>\delta {\Gamma}^s_{ia} - \frac{1}{2} \, g_{ij} \, \nabla_b \delta g^{ab} \, \nabla_a \delta g^{ij} + \frac{1}{4} \, g^{ab} \, g_{mn} \, g_{ij} \, \nabla_b \delta g^{mn}\, \nabla_a \delta g^{ij} \\
& \>\>\>\>\>\>\>\>\>\>\>\>\>\>\>\>\>\>\>\>\>\>\>\>\>\>\>\>\>\>   + \frac{1}{2} \left( R \, Y_{abmn}  -   {R}_{ab} \,g_{mn}\right) \delta g^{ab} \> \delta g^{mn} \biggr] \sqrt{g} \> d^3 y .
\end{aligned}
\end{equation}

\noindent I again change index labels and insert Kronecker deltas to simplify further:
\begin{equation}\label{A-3EHVar4A4}
\begin{aligned}
\Delta S_{EH} 
& = \delta S_{EH} + \int_{\Sigma} \biggl[  - \nabla_s \delta g^{ab} \, \delta {\Gamma}^s_{ba} - g^{ab} \, \delta {\Gamma}^i_{bs} \>\delta {\Gamma}^s_{ia} - \frac{1}{2} \, g_{mn} \, \delta_i^b \, \delta_j^a \, \nabla_i \delta g^{ab} \, \nabla_j \delta g^{mn} + \frac{1}{4} \, g^{ij} \, g_{mn} \, g_{ab} \, \nabla_j \delta g^{mn}\, \nabla_i \delta g^{ab} \\
& \>\>\>\>\>\>\>\>\>\>\>\>\>\>\>\>\>\>\>\>\>\>\>\>\>\>\>\>\>\>   + \frac{1}{2} \left( R \, Y_{abmn}  -   {R}_{ab} \,g_{mn}\right) \delta g^{ab} \> \delta g^{mn} \biggr] \sqrt{g} \> d^3 y .
\end{aligned}
\end{equation}

\noindent I now define the following two scalar quantities:
\begin{equation} \label{A-ConnectionScalars}
\begin{aligned}
A_1 &= \nabla_s \delta g^{ab} \, \delta {\Gamma}^s_{ba} \\
A_2 &= g^{ab} \, \delta {\Gamma}^i_{bs} \>\delta {\Gamma}^s_{ia} ,
\end{aligned}
\end{equation}

\noindent so that the variation of the action becomes [after collecting terms in (\ref{A-3EHVar4A4})]:
\begin{equation}\label{A-3EHVar4B}
\begin{aligned}
\Delta S_{EH} & = \delta S_{EH} + \int_{\Sigma} \biggl[  - (A_1 + A_2) + \left(\frac{1}{4} \, g^{ij} \, g_{mn} \, g_{ab} - \frac{1}{2} \, g_{mn} \, \delta_i^b \, \delta_j^a \right) \nabla_i \delta g^{ab} \, \nabla_j \delta g^{mn} \\
& \>\>\>\>\>\>\>\>\>\>\>\>\>\>\>\>\>\>\>\>\>\>\>\>\>\>\>\>\>\>   + \frac{1}{2} \left( R \, Y_{abmn}  -   {R}_{ab} \,g_{mn}\right) \delta g^{ab} \> \delta g^{mn} \biggr] \sqrt{g} \> d^3 y .
\end{aligned}
\end{equation}

I now work out explicit expressions for $A_1$ and $A_2$, using the expressions Eq. (\ref{SFD-ChristoffelSymbolVariationC1}). $A_1$ is relatively simple to work out:
\begin{equation}\label{SFD-Quantity1}
\begin{aligned}
A_1 	&= \nabla_s \delta g^{ab} \, \delta {\Gamma}^s_{ba} \\
&=\frac{1}{2} \nabla_s \delta g^{ab} \, (g_{mb} \, g_{na} \, g^{sk}\,  \nabla_k \delta g^{mn} - g_{na} \, \nabla_b \delta g^{sn} - g_{mb} \, \nabla_a \delta g^{ms})\\
		&=\frac{1}{2} (g_{mb} \, g_{na} \, g^{ij}\, \nabla_i \delta g^{ab} \,  \nabla_j \delta g^{mn} - 2 \, g_{na} \, \nabla_m \delta g^{ab} \,  \nabla_b \delta g^{mn}).
\end{aligned}
\end{equation}

\noindent I perform additional index relabelings and insert Kronecker deltas to obtain
\begin{equation}\label{SFD-Quantity1B}
\begin{aligned}
A_1
&= \frac{1}{2} (g_{mb} \, g_{na} \, g^{ij}  - 2 \, g_{na} \, \delta_m^i \, \delta_b^j )\nabla_i \delta g^{ab} \,  \nabla_j \delta g^{mn}.
\end{aligned}
\end{equation}

\noindent The computation of $A_2$ is more involved [again, I use Eq. (\ref{SFD-ChristoffelSymbolVariationC1})]:
\begin{equation}\label{SFD-Quantity2a}
\begin{aligned}
A_2 &= g^{ab} \, \delta {\Gamma}^i_{bs} \>\delta {\Gamma}^s_{ia} \\
& = \frac{1}{4} \, g^{ab} \, (g_{mb} \, g_{ns} \, g^{ik}\,  \nabla_k \delta g^{mn} - g_{ns} \, \nabla_b \delta g^{in} - g_{mb} \, \nabla_s \delta g^{mi}) \,  (g_{pi} \, g_{qa} \, g^{sk}\,  \nabla_k \delta g^{pq} - g_{qa} \, \nabla_i \delta g^{sq} - g_{pi} \, \nabla_a \delta g^{ps})\\
& = \frac{1}{4} \, g^{ab} \, \biggl(g_{mb} \, g_{ns} \, g^{ir}\,  \nabla_r \delta g^{mn} \, g_{pi} \, g_{qa} \, g^{sk}\,  \nabla_k \delta g^{pq} - g_{mb} \, g_{ns} \, g^{ir}\,  \nabla_r \delta g^{mn} \, g_{qa} \, \nabla_i \delta g^{sq} - g_{mb} \, g_{ns} \, g^{ir}\,  \nabla_r \delta g^{mn} \,g_{pi} \, \nabla_a \delta g^{ps} \\
& \>\>\>\>\>\>\>\>\>\>\>\>\>\>\>\>\>\>\>\> -  g_{ns} \, \nabla_b \delta g^{in} \, g_{pi} \, g_{qa} \, g^{sk}\,  \nabla_k \delta g^{pq} +  g_{ns} \, \nabla_b \delta g^{in} \, g_{qa} \, \nabla_i \delta g^{sq} +  g_{ns} \, \nabla_b \delta g^{in} \, g_{pi} \, \nabla_a \delta g^{ps}\\
& \>\>\>\>\>\>\>\>\>\>\>\>\>\>\>\>\>\>\>\> -  g_{mb} \, \nabla_s \delta g^{mi} \, g_{pi} \, g_{qa} \, g^{sk}\,  \nabla_k \delta g^{pq} +  g_{mb} \, \nabla_s \delta g^{mi} \, g_{qa} \, \nabla_i \delta g^{sq} +  g_{mb} \, \nabla_s \delta g^{mi} \, g_{pi} \, \nabla_a \delta g^{ps} \biggr).
\end{aligned}
\end{equation}

\noindent After performing some contractions and index relabeling, the above becomes
\begin{equation}\label{SFD-Quantity2b}
\begin{aligned}
A_2 
& = \frac{1}{4} \, \biggl(g_{mb}  \, \nabla_a \delta g^{mn} \,  \nabla_n \delta g^{ab} - \underline{g_{mb} \, g_{ns} \, g^{ir}\,  \nabla_r \delta g^{mn} \, \nabla_i \delta g^{sb}} - \underline{g_{ns} \, \nabla_r \delta g^{an}  \, \nabla_a \delta g^{rs}} \\
& \>\>\>\>\>\>\>\>\>\>\>\>\>\>\> -  \underline{g_{pi} \, \nabla_b \delta g^{in} \,  \nabla_n \delta g^{pb}} + \underline{g_{ns} \, \nabla_r \delta g^{an} \, \nabla_a \delta g^{sr}} +  \underline{ g_{bm} \, g_{ns} \, g^{ir}  \, \nabla_r \delta g^{mn} \, \nabla_i \delta g^{bs}}\\
& \>\>\>\>\>\>\>\>\>\>\>\>\>\>\> - g_{mb} \, g_{pi}  \, g^{sk}\, \nabla_s \delta g^{mi} \,   \nabla_k \delta g^{pb} + g_{mb} \, \nabla_a \delta g^{mn} \, \nabla_n \delta g^{ab} + \underline{g_{pi} \,\nabla_b \delta g^{ni} \,  \nabla_n \delta g^{pb}} \biggr).
\end{aligned}
\end{equation}

\noindent The underlined terms cancel and I obtain the following expression:
\begin{equation}\label{SFD-Quantity2bb}
\begin{aligned}
A_2 & = \frac{1}{4} \, ( 2 g_{mb} \, \nabla_a \delta g^{mn} \, \nabla_n \delta g^{ab} - g_{mb} \, g_{pi}  \, g^{sk}\, \nabla_s \delta g^{mi} \, \nabla_k \delta g^{pb} )\\
	& = \frac{1}{4} \, ( 2 g_{mb} \, \delta_a^j \, \delta_n^i \, \nabla_j \delta g^{mn} \, \nabla_i \delta g^{ab} - g_{mb} \, g_{an}  \, g^{ij}\, \nabla_j \delta g^{mn} \, \nabla_i \delta g^{ab} )\\
& = \frac{1}{4} \, ( 2 g_{mb} \, \delta_a^j \, \delta_n^i - g_{mb} \, g_{an}  \, g^{ij} )\nabla_i \delta g^{ab} \, \nabla_j \delta g^{mn}.
\end{aligned}
\end{equation}

I now insert Eqs. (\ref{SFD-Quantity1B}) and (\ref{SFD-Quantity2bb}) into Eq. (\ref{A-3EHVar4B}) to obtain the following expression for $\Delta S_{EH}$, which I simplify as
\begin{equation}\label{A-3EHVar4B1}
\begin{aligned}
\Delta S_{EH} & = \delta S_{EH} + \int_{\Sigma} \biggl[ \left( g_{na} \, \delta_m^i \, \delta_b^j - \frac{1}{2} \, g_{mb} \, g_{na} \, g^{ij}  - \frac{1}{2} \, g_{mb} \, \delta_a^j \, \delta_n^i + \frac{1}{4} \, g_{mb} \, g_{an}  \, g^{ij}\right) \nabla_i \delta g^{ab} \, \nabla_j \delta g^{mn} \\
& \>\>\>\>\>\>\>\>\>\>\>\>\>\>\>\>\>\>\>\>\>\>\>\>\>\>\>\>\>\>  + \left(\frac{1}{4} \, g^{ij} \, g_{mn} \, g_{ab} - \frac{1}{2} \, g_{mn} \, \delta^i_b \, \delta^j_a \right) \nabla_i \delta g^{ab} \, \nabla_j \delta g^{mn}  + \frac{1}{2} \left( R \, Y_{abmn}  -   {R}_{ab} \,g_{mn}\right) \delta g^{ab} \> \delta g^{mn} \biggr] \sqrt{g} \> d^3 y \\
& = \delta S_{EH} + \int_{\Sigma} \biggl[ \frac{1}{4}\left(4\, g_{na} \, \delta_m^i \, \delta_b^j - 2 \, g_{mb} \, \delta_a^j \, \delta_n^i - g_{mb} \, g_{an}  \, g^{ij} + g^{ij} \, g_{mn} \, g_{ab} - 2 \, g_{mn} \, \delta^i_b \, \delta^j_a \right) \nabla_i \delta g^{ab} \, \nabla_j \delta g^{mn} \\
& \>\>\>\>\>\>\>\>\>\>\>\>\>\>\>\>\> + \frac{1}{2} \left( R \, Y_{abmn}  -   {R}_{ab} \,g_{mn}\right) \delta g^{ab} \> \delta g^{mn} \biggr] \sqrt{g} \> d^3 y .
\end{aligned}
\end{equation}

\noindent Finally, I write
\begin{equation}\label{A-3EHVar4C}
\begin{aligned}
\Delta S_{EH}&= \delta S_{EH} + \int_{\Sigma} \biggl[Z_{abmn}^{ij} \nabla_i \delta g^{ab} \, \nabla_j \delta g^{mn} + \frac{1}{2} \left( R \, Y_{abmn}  -   {R}_{ab} \,g_{mn}\right) \delta g^{ab} \> \delta g^{mn} \biggr] \sqrt{g} \> d^3 y ,
\end{aligned}
\end{equation}

\noindent where I define the following quantity:
\begin{equation} \label{A-ZTensor}
\begin{aligned}
Z_{abmn}^{ij} &:=\frac{1}{4}\biggl(4\, g_{na} \, \delta_m^i \, \delta_b^j - 2 \, g_{mb} \, \delta_a^j \, \delta_n^i - g_{mb} \, g_{an}  \, g^{ij} + g^{ij} \, g_{mn} \, g_{ab} - 2 \, g_{mn} \, \delta^i_b \, \delta^j_a  \biggr) .
\end{aligned}
\end{equation}

\bibliography{bibStd,bibGRQG,bibGRADM,bibSptop,bibOurs,bibSFD}

\end{document}